%Paper: hep-th/9412241
%From: volovich@arevol.mian.su (Volovich Igor)
%Date: Sat, 31 Dec 94 21:43:12 +0300 (MSK)

\magnification\magstep1
$$~~~~~~~~~~~~~~~~~~~~~~~~~~~~~~~~~~~~~~~~~~~~~~~~~~~~~~~~~~~~~~
{}~~~~~~~~~~~~~~~~~~~~~~~~~~~~~CVV-198-94$$
\bigskip
\centerline{\bf Non--Commutative (Quantum) Probability, }
\smallskip
\centerline{\bf Master Fields and Stochastic Bosonization }
\bigskip\bigskip\bigskip
\centerline {\bf L. Accardi\qquad Y.G.
Lu\footnote{$^{(*)}$}{\rm Dipartimento di Matematica, Universit\`a di Bari.}
\qquad I. Volovich\footnote{$^{(**)}$}{\rm On leave absence from Steklov
Mathem.Inst, Vavilov St.42,GSP--1,117966, Moscow.}}
\bigskip\bigskip
\centerline {\bf Centro V. Volterra}\smallskip
\centerline{\bf Universit\`a di Roma Tor Vergata}\smallskip
\centerline{\bf via di Tor Vergata, 00133 -- Roma}\smallskip
\centerline{e--Mail volterra@mat.utovrm.it}
\bigskip\bigskip\bigskip

\noindent{\it Abstract\/}. In this report we discuss some results of
non--commutative (quantum) probability theory relating the various
notions of statistical independence and the associated quantum central
limit theorems to different aspects of mathematics and physics
including: $q$--deformed and free central limit theorems; the description
of the master (i.e. central limit) field in matrix models along the recent
Singer suggestion to relate it to Voiculescu's results on the freeness
of the large $N$ limit of random matrices; quantum stochastic differential
equations for the gauge master field in QCD; the theory of  stochastic
limits of quantum fields and its applications to stochastic bosonization of
Fermi fields in any dimensions; new structures in QED such as a nonlinear
modification of the Wigner semicircle law and  the interacting
Fock space: a natural explicit example of a self--interacting quantum field
which exhibits the non crossing diagrams of the Wigner semicircle law.

\bigskip
(1.) Introduction

(2.) Algebraic (quantum, non--commutative) probability

(3.) Independence

(4.) Fock spaces (Boson, Fermion, Free) and their deformations

(5.) Gaussianity (classical, Boson, Fermion, Free, $\ldots $)

(6.) Algebraic laws of large numbers and central limit theorems: master fields

(7.) Large $N$ limit for matrix models and asymptotic freeness

(8.) Quantum stochastic differential equations

(9.) Stochastic  limits and anisotropic asymptotics

(10.) Stochastic bosonization

(11.) The interacting Fock module and QED
\bigskip

\beginsection{(1.) Introduction  }

\bigskip
In this report we discuss some basic results of non-commutative
(quantum) probability theory and their applications to large $N$
limit and stochastic  limit in quantum field theory.
1/N expansion is probably one of the most powerful methods
in investigation of essential non-linear effects
in quantum field theory and statistical physics.
There are two main types of models admitting the large N expansion.
These are so called vector and matrix models.
Matrix models are more difficult for investigation. 1/N expansion was
used  to describe phase transition of N component spin models [St,BrZ].
It was shown [Ar76] that for vector models the using
$1/N$ expansion permits to prove the renormalizability
of 3 dim $\sigma$-model which is nonrenormalizable
in the standard perturbation theory.
The large $N$ limit for vector models can be
interpreted as a central limit theorem
from the point of view of probability theory [AAV93].
For matrix model the  large N colour expansion introduced  by 't Hooft
[tHo74] is one of most promising technique for arriving at
an analytical understanding
of long distance
properties of non abelian gauge theories. The major obstacle
to realization of this program is the fact that it has been
impossible up to now
to compute the leading term in this expansion, i.e.
the sum of the planar graphs in the closed form
unless one works in zero or one dimension [BIPZ].
Many matrix models simplify greatly in the limit of large
$N$ and it was suggested that the limiting behaviour
of a model can be described in terms of so called
the {\it master field} [Wit,MM, Ha80, Are81].
An  equation for the Wilson loop functional
in the large N limit was obtained by Makeenko and Migdal [MM].

In what follows we will explain that  the master field for the simplest
vector model is nothing but the operator of
coordinate of the standard quantum mechanical
harmonic oscillator. Recently Singer [Sin94], based on Voiculescu's
works, has shown that the master field for matrix models
can be described  in terms of free random variables satisfying
free commutation relations.
The free commutation relations is a q-deformation of canonical commutation
relations for $q=0$.
Non-commutative (quantum) probability theory considers
central limit theorems for $q$-deformed commutation relations
and provides an unified approach to both vector and matrix models.

The simplest classical central
limit theorem for Gaussian random variables reads
$$
\lim_{N\to\infty}\langle S^k_N\rangle_N
=\int_R \lambda^k\varrho(\lambda)d\lambda
\eqno(1.1)$$
where $k=1,2,...$,$$S_N={1\over \sqrt{ N}}\sum_{i=1}^N x_i,$$
$$
\langle f(x)\rangle_N={1\over Z_N}\int_{R^N} f(x)
e^{-{x^2\over 2}}dx \eqno(1.2)
$$
$$
Z_N=\int_{R^N} e^{-{x^2\over 2}}dx=(2\pi)^{{N\over 2}}\ ,\qquad
x=(x_1,...,x_n)
$$
The function $\varrho(\lambda)$ is the Gaussian density,
$$
\varrho(\lambda)=
{1\over\sqrt{2\pi}}\,e^{-{\lambda^2\over 2}}\eqno(1.3)
$$
Below we will use also other notations for the
expectation value,
$$
\langle f\rangle =Ef=\varphi (f)
$$

One can rewrite (1.1) as follows. Let us introduce an operator
$$
Q=a^* +a\eqno(1.4)
$$
where $a$ and $a^*$ are usual annihilation and creation operators  satisfying
$$
[a,a^*]=1
\eqno(1.5)
$$
Then one has
$$\int_R \lambda^k\varrho(\lambda)d\lambda=
(\Phi,Q^k\Phi)
\eqno(1.6)$$
The vector $\Phi$ in (1.6) is the vacuum
vector in the Fock space for the quantum harmonic
oscillator, $a\Phi=0$. As it is well known
this Fock space is isomorphic to $L^2(R)$.

We can interpret the position operator $Q$ as a
{\it master field\/} (a {\it random variable\/} in the sense of quantum
probability).
So, the master field for simplest vector model is the operator
of  coordinate for harmonic oscillator,
$$
\lim_{N\to\infty}\langle S^k_N\rangle_N=
(\Phi,Q^k\Phi).
\eqno(1.7)$$

If the covariance of our random variables
is not 1 but  $\hbar$,
$$
\langle x_i^2\rangle_N=\hbar
$$
then instead of (1.5) one gets
$$
[a,a^*]=\hbar\eqno(1.8).
$$
Now, the position operator gives (roughly speaking)
{\it one half\/} of the quantum observables: the other half being given
by the momentum operator $P=(a-a^+)/i$. In this sense
one can say that we can get {\it half\/} quantum mechanics
(more exactly, an abelian subalgebra
of the quantum mechanical algebra generated
by the operators of momenta and position)
as a limit in classical probability
theory. In order to obtain the {\it  other half\/} the classical central
limit theorems are not sufficient and one must look at the quantum
central limit theorems. This was done in the late seventies.

For a non-Gaussian family of independent
random variables the central limit theorem
can be formulated in a form similar
to (1.1), (1.7) but now
$$
S_N={1\over\sqrt{ D_{ N}}}\sum_{i=1}^N x_i\eqno(1.9)
$$
where the covariance is $D_N=\langle S^2_N\rangle_N$
and the exponential function in (1.2) can have
non-quadratic terms. If the limit in (1.7)
is non-Gaussian, then the
master field $Q$
will be a non-linear function of $a$ and $a^*$,
$Q=Q(a,a^*)$.

Now let us discuss matrix models.
Let $A=(A_{ij})_{i,j=1}^N$ be an ensemble of real symmetric
$N\times N$ matrices with distribution given by
$$
\langle f(A)\rangle_N={1\over Z_N}\int f(A)
e^{-NS(A)}\prod_{i\leq j}dA_{ij},\eqno(1.10)$$
$$Z_N=\int e^{-NS(A)}\prod_{i\leq j}dA_{ij}$$
where the {\it action\/} $S(A)$ is quadratic,
$$
S(A)={1\over 2}trA^2\eqno(1.11)
$$
It follows from the works of Wigner [Wig] and Voiculescu [Voi92]
that the limits
$$
\lim_{N\to\infty}\langle {1\over N}tr(A^k)\rangle_N=
(\Phi,Q^k\Phi)\qquad;\qquad k=1,2,\dots\eqno(1.12)$$
exists. The limiting object $Q$ in (1.1) is a quantum random variable
(in the sense of Section (2.) below) which
is called
{\it the master field}. It is defined as
$$
Q=a^* +a \eqno(1.13)$$
where $a^*$ and $a$ are {\it free}
creation and annihilation operators, i.e.
they do not satisfy the usual canonical
commutation relations but the following
$$
aa^* =1 \eqno(1.14)
$$
The vector $\Phi$ in (1.1) is a {\it vacuum vector\/}
in a {\it free} Fock space,
$$
a\Phi =0
$$
The free Fock space is constructed starting from $a^*$ and $\Phi$,
by the usual procedure but only
by using the operators $a$ and $a^*$
satisfying (1.14) (cf. Section (4.) for more details).
The expectation value in (1.12) is
known to be
$$
(\Phi,Q^k \Phi)=\int_R \lambda^k
w(\lambda)d\lambda \eqno(1.15)
$$
where $w(\lambda)$ is the Wigner semicircle
density,
$$
w(\lambda)={1\over 2\pi}\sqrt {4-\lambda^2}\qquad,
\ \ |\lambda|\leq 2 \eqno(1.16)
$$
and $w(\lambda)=0$ for $|\lambda|\geq 2$.
The Wigner semicircle distribution in noncommutative probability
plays the role of the Gaussian distribution in classical
probability (cf. Section 7 for a substantiation of this statement).

Generalizations of the usual commutation (or anticommutation) relations
have been studied, in relation to quantum groups (cfr., for example,
[AV91, APVV]) and to quantum probability (cf. [Sch\"u93]).
One can consider
the relation (1.14) as a limiting case $(q=0)$
of $q$-deformed commutation relations,
$$
aa^*-qaa^*=\hbar  \eqno(1.17)
$$
which reduce to the usual quantum
mechanical oscillator and classical probability
theory when $q=1$. From this point of view a $q$--deformation ($q=0$)of
canonical commutation relations corresponds
to a transition from the vector models to matrix models in the large
N limit.

If one considers a more general action of the form
$$S(A)=S_0(A)+S_I(A)$$
where $S_0(A)$ has the form (1.11) and
$S_I(A)$ is a more general nonlinear
functional, then expanding the term $\exp S_I(A)$ in power series and
exchanging formally
the summation and the limit in (1.12), one still is
lead to consider limits
of the form (1.12). For this reason it is quite
natural to conjecture,
that the Voiculescu results should find a natural
application in the large $N$ expansion of quantum field models.
In particular
in QCD would be important to describe the large
$N$ behaviour. Recently Singer [Sin94] has suggested to
use Voiculescu's results on free random variables
in non-commutative probability theory to describe the master
field and he considered the master field for 2 dimensional QCD.
This was further discussed in [Do94,GG94].

Singer's conjecture is also supported by
another result, obtained a couple of
years ago [AcLu91], [AcLu93a], [AcLu93b])
which shows a surprising emergence
of the semi--circle
diagrams (not of the semicircle law but a nonlinear
modification of it) in
the stochastic limit of quantum electrodynamics.
Since QCD is a generalization of QED it
is natural to expect that these diagrams should play a role also in QCD
(cf. Section (11.) below for an outline of the QED application and
[AcLuVo94a]  for a preliminary result concerning QCD: the latter paper,
although written independently and based on completely different
techniques goes in the same direction of the recent work [Wil94].

The master field as a limiting
quantum random process there appears not only in the large N
limit but also in the so colled stochastic limit
of quantum field theory.
In [AcLuVo93] we have found the following
direct relation between quantum theory
in real time and stochastic processes:
$$\lim_{\lambda\to0}\lambda A\left({t\over\lambda^2}\,,\ k\right)=B(t,k)
\eqno(1.18)$$
where $A(t,k)$ is a free dynamical evolution of a usual (Boson or Fermion)
annihilation
operator $a(k)$ in a Fock space ${\cal F}$, i.e.
$$A(t,k)=e^{itE(k)}a(k)$$
$k$ is a momentum variable and $B(t,k)$ is a Boson or Fermion quantum
field acting in another Fock space ${\cal H}$ and
satisfying the canonical commutation relations:
$$[B(t,k),B(t',k')]_\pm=2\pi\delta(t-t')\delta(k-k')
\delta(E(k))\eqno(1.19)$$

A (Boson or Fermion) Fock field satisfying commutation relations of the
form (1.19) is called a {\it quantum white noise\/}
[Ac Fri Lu1] and is a
prototype example of quantum stochastic process. In the context
of this our paper one can call it also the master field.
In terms of the field operators $\phi(t,x)$ one
can rewrite (1.18) as
$$\lim_{\lambda\to0}\lambda\phi\left({t\over\lambda^2}\,,\
x\right)=W(t,x)\eqno(1.20)$$
and one can prove that the vacuum correlations of
$W(t,x)$ coincide with those of a {\it classical Brownian motion\/}.
The relation (1.18) can be interpreted as a quantum central limit
theorem with $\lambda$ playing the role of ${1\over {\sqrt N}}$.

The scaling $t\mapsto t/\lambda^2$ has its origins
in the early attempts by Pauli,
Friedrichs and Van Hove to deduce a {\it
master equation\/}.

In a series of papers [AcLu85],[AcLu93]
it has been shown that the fact that,
under the
scaling $t\to t/\lambda^2$, quantum fields converge to quantum
Brownian motions and
the Heisenberg equation to a quantum Langevin
equation, is a rather
universal phenomenon as shown in a multiplicity of
quantum systems, involving the basic physical interactions.

This kind of limit and
 the set of mathematical techniques developed to
established it,
was called in [AcLuVo93] {\it the stochastic limit of quantum field
theory\/}. We call it also the $1+3$ asymptotical expansion to
distinguish it from analogous scalings generalizing the present one in
the natural direction of rescaling other variables beyond (or instead
of) time: space, energy particle density,...

A generalization of (1.18) to a pair of
{\it Fermi operators\/} $A,A^+$ has been obtained in [AcLuVo94]
and it can be
described by the formula
$$\lim_{\lambda\to0}
\lambda A\left({t\over\lambda^2}\,,\ k_1\right)A
\left({t\over\lambda^2}\,,\ k_2\right)=B(t,k_1,k_2)\eqno(1.21)$$
The remarkable property of the formula (1.21) is
that while the $A(t,k)$
are {\bf Fermion}
annihilation operators the limit field $B(t,k_1,k_2)$ is a
{\bf Boson}
annihilation operator which satisfies with its hermitian
conjugate, the canonical bosonic commutation relations. For this
reasons we call the
formula (1.21): {\it stochastic bosonization\/}.

The bosonization of Fermions is well known in $1+1$ dimensions, in
particular in the Thirring--Luttinger model see, for example [Wh,
SV] and in string theory [GSW]. The stochastic bosonization (1.21)
takes place in the real $1+3$
dimensional space--time and in fact in any
dimension.
For previous discussion of
bosonization in higher dimensions cf. [Lut],
[Hal].

In particular in quantum chromodynamics one can think of the
Fermi--operators $A(t,k)$ as corresponding to quarks then the bosonic
operator $B(t,k_1,k_2)$ can be considered as describing a meson (cf.
[AcLuVo94b] for a preliminary approach to QCD in this spirit).

An extension of the notion of the stochastic limit
is the notion of {\it  anisotropic asymptotics}[AV94].
Anisotropic asymptotics
describe behaviour of correlation functions when some components of
coordinates are large as compare with others components. It is
occurred that (2+2) anisotropic asymptotics for 4-points functions
are related with the well known Regge regime of the
scattering amplitudes.
Let us explane  what we mean by  anisotropic
asymptotics.
If  $x^{\mu}$ are space-time coordinates, one denotes
$x^{\mu}=(y^{\alpha}, z^{i})$, where $\alpha =0,1$, $i=2,3$ for 2+2
decomposition and  $\alpha =0$,
$i=1,2,3$ for 1+3 decomposition. Anisotropic asymptotics mean
the evaluation of asymptotics of correlation
functions for a field  $\Phi (\lambda y,z)$
$$
<\Phi _{i_{1}}
(\lambda y_{1},z_{1})\Phi _{i_{2}}(\lambda y_{2},z_{2})...
\Phi _{i_{n}}(\lambda y_{n},z_{n})>
$$
when $\lambda\to 0$ (or $\lambda\to \infty$), i.e.
when some of variables are much larger than others.

In the following section we shall describe some results of quantum
probability relevant for the present exposition (for a more detailed
review cf. [Ac90]) and then we consider a
generalization of the relations of (1.7) and (1.12) to the case of
several matrix random variables, to the non--Gaussian case and to the
case of $q$--deformed algebras of the CCR.\bigskip\medskip

\beginsection{\bf (2.) Algebraic (quantum,
non--commutative) probability }

\bigskip

A positive linear
$*$-functional $\varphi $ on a $*$--algebra ${\cal A}$
(i.e. a
complex algebra with an involution denoted $*$ and a unit denoted $1$)
 such that $\varphi (1) = 1 $ is called a {\bf state}.
An {\bf algebraic
probability space} is a pair $\{{\cal A},\ \varphi\}$ where
${\cal A}$ is a $^*$--algebra and $\varphi$ is a state on ${\cal A}$.

If ${\cal A}$ is a commutative algebra we speak of a {\bf classical
probability space}; if ${\cal A}$ is noncommutative of a {\bf quantum
probability space}.

If ${\cal A}$ is a commutative separable von Neumann algebra, then it is
isomorphic
to the algebra
$L^\infty (\Omega , {\cal F},P) $ where $(\Omega ,{\cal F},P)$
is a probability space
in the usual sense ($\Omega $ is a set, ${\cal F}$
a $\sigma $--algebra of subsets of $\Omega $, $P$ a positive measure
with $P(\Omega ) =1 $).
Moreover, in this isomorphism, the state $\varphi$
becomes the usual integral with respect to $P$, i.e.
$\int_\Omega \ (\cdot) \ d P $.
% If ${\cal A}$ is commutative but not a von
% Neumann
algebra, then, with the Gelfand--Naimark--Segal (GNS) construction
with respect to $\varphi$.
% one associates a commutative von Neumann algebra to the algebraic
% probability space $\{{\cal A},\ \varphi\}$
Thus classical probability is naturally included in algebraic probability.
In fact one can prove that there is an isomorphism, in the sense of
categories, between classical probability spaces in the algebraic sense
and probability spaces in the sense of classical probability theory,
which form a category
with morphisms defined by measure preserving measurable
transformations defined up to sets of zero probability.

%%%%%%
An investigation of such and more general
relations as (1.1)
and (1.5) is one  of subjects of non-commutative
(quantum) probability theory.

The oldest example of quantum
probability space is the pair $\{{\cal B}({\cal H}), \varphi \}$ where
${\cal H}$ is a Hilbert space and $\varphi $ a state on the algebra
${\cal B}({\cal H})$ of all bounded operators on ${\cal H}$. Typical
examples of states are $ \varphi (a) = <\Phi , a \Phi >$ where $\Phi $
is a vector in ${\cal H}$ (e.g. Fock vacuum) or
$ \varphi(a) = Tr (wa)$, where $w$ is a density matrix (e.g. an
equilibrium finite temperature state).

A {\it random variable\/} with probability space $\{{\cal A},\varphi\}$
is an element of ${\cal A}$.

A {\it stochastic process\/}
is family $(X_t)$, of random variables, indexed by
an arbitrary set $T$.
Given a stochastic process $X\equiv (X_t)$, one can form
the polynomial $*$--
algebra ${\cal P} (X) $, which is a $*$--subalgebra of
${\cal A}$. The restriction of the state $ \varphi $ on this algebra
gives a state $ \varphi _X $ on ${\cal P} (X) $ which is called the
{\bf distribution} of the process $X$. If $T= \{ 1, \ldots , n\}$ is a
finite set, the
state $ \varphi _X $ is called the {\bf joint distribution}
of the random variables $\{ X_1, \ldots , X_n\}$. If $T$ consists of a
single element, then one speaks of the {\bf distribution} of the random
variable $X$.

It is known that
the free $*$--algebra $ {\cal C} < \tilde X_t \ : \ t\in T > $
generated by the (noncommuting) indeterminates $\tilde X_t \ (t\in T ) $
has the following universal property: for any $*$--algebra ${\cal P} $
and for any family $\{X_t \ : \ t\in T \} $ of its elements, there
exists a unique homomorphism
$ \alpha : {\cal C} < \tilde X_t \ : \ t\in T > \to {\cal P} $,
characterized by the property:
$$ \alpha (\tilde X_t ) = X_t \qquad ; \ \forall t\in T $$
Using this $\alpha $ one can {\it pull back } the state
$ \varphi $ on ${\cal A}$ to the state
$ \tilde \varphi := \varphi \circ \alpha $ on the free algebra and,
because of the universal property, the algebraic probability space
$$ \{ {\cal C} < \tilde X_t \ : \ t\in T > , \tilde \varphi \} \eqno(2.1) $$
contains all the informations on the stochastic process $X$. Summing up:
{\bf to give an algebraic stochastic process indexed by a set $T$, is
the same as giving a probability space of the form (2.1)}.

This {\it pull back\/} is useful because {\it it depends only on the
index set $T$\/}. This implies that, given a sequence (or, more
generally, a net) of stochastic processes
$$X^{(n)}\equiv\{{\cal A}^{(n)},X^{(n)}_t,t\in T,\ \varphi^{(n)}\}\
;\quad n=1,2,\dots\eqno(2.2)$$
one can define a sequence (net) $(\tilde\varphi^{(n)})$ of states {\bf
on the same algebra} ${\cal C}\langle\tilde X_t:t\in T\rangle$ and
therefore it makes sense to speak of convergence of $(\varphi^{(n)})$ to
a limiting state $\varphi^{(\infty)}$ on ${\cal C}\langle\tilde X_t:t\in
T\rangle$. By the above remark, the algebraic probability space $\{{\cal
C}\langle\tilde X_t:t\in T\rangle$, $\varphi^{(\infty)}\}$ uniquely
defines an algebraic stochastic process which is called {\bf the limit
in law} (or in distribution) of the sequence of processes $X^{(n)}$
given by (2.2). This notion of convergence in law was pioneered by von
Waldenfels [vWI88] and used by Voiculescu [Voi92].
For results such as the law of large
numbers and the central limit theorem (cf. Section 8.) below) this notion
is sufficient, but for more interesting applications to physics, a more
sophisticated notion (which turns out to be nearer to the spirit of
traditional quantum theory) is required (cf. Section (12.) below).

The notion of convergence in law plays a crucial role in the stochastic
limit of quantum field theory (cf. Section () below) and it is likely to
play a similar role in any constructive developement of quantum field
theory.

In fact it is known (von Neumann--Friedrichs--Haag
phenomenon) that an interacting field cannot live on the same Hilbert
space of its associated free field. Therefore, if we want to
approximate, in some sense, the interacting correlations by local
perturbations of free correlations, then we cannot use the standard
operator topologies, which do not allow to {\it go out\/} from the
original space.

Now let ${\cal F}$ be the (GNS) space of this probability space, $\Phi_F$
the cyclic vector and $\pi$ the Gelfand--Naimark--Segal
GNS representation of
${\cal C} < \tilde X_t \ : \ t\in T > $ in the linear operators on
${\cal F}$. Denoting ${\cal H}_1 $ the Hilbert subspace of ${\cal F}$
generated by the vectors $\{ \pi (\tilde X_t ) \cdot \Phi_F \ : \ t\in T
\}$ (the 1--particle space),
we can identify ${\cal F}$ to the full Fock space $\Gamma ({\cal H}_1 )$
over the {\it one particle space} $ {\cal H}_1 $ with a unitary
isomorphism that maps $\Phi _F$ to the vacuum vector $\Phi $ and
intertwines the GNS representation of ${\cal C} < \tilde X_t \ :\ t\in T >$
with tensor multiplication (these two
properties uniquely determine the unitary).

The {\it free (or full) Fock space\/} ${\cal F}(H)$ over a
pre--Hilbert space $H$ is just the tensor algebra over $H$, i.e.
$${\cal F}(H) :
=\oplus_{n\geq0}(\otimes H)^n={\bf C}\cdot \Phi \oplus H\oplus
H\otimes H\oplus\dots$$
where $\Phi =1$ is the vacuum vector. The Boson (resp. Fermion) Fock space
differs from it because the tensor product in the $n$--particle space is
symmetrized (resp. antisymmetrized).

The creation and annihilation operators in the full Fock space are defined by
$$A(f)\Phi =0$$
$$A(f)g_1\otimes g_2\otimes\dots\otimes g_n=\langle f,g_1\rangle
g_2\otimes\dots\otimes g_n$$
$$A^+(f)g_1\otimes\dots\otimes g_n=f\otimes g_1\otimes\dots\otimes g_n$$
Here $f$, $g_1,\dots,g_n$ are elements from $H$. From these definitions
it follows that
$$A(f)A^+(g)=\langle f,g\rangle\eqno(2.3)$$
Operators $A(f)$ and $A^+(f)$ are bounded operators in the Hilbert space
${\cal F}(H)$.

Notice that, in the above mentioned identification between the GNS space of
the free algebra with the full Fock space, the GNS representation $\pi
(X_t)$ of a random variable $X_t$ becomes the creation operator
$a^+(X_t)$.

Summing up: since the full Fock algebra is a universal object for
stochastic processes (in the sense explained above), it follows that it
is a natural programme to study subalgebras of this algebra and states
on them. The symmetric (Boson) and antisymmetric (Fermion) algebra and
corresponding classes of states have been widely studied.
Another interesting subalgebra is the so--called {\it Toeplitz algebra},
which is the norm closure of the algebra generated by the creation and
annihilation operators, plus the identity (in the identification of
$\Gamma({\cal H}_1 )$ with ${\cal H}_F$
this is simply the norm closure of the GNS representation of
${\cal C} < \tilde X_t \ : \ t\in T > $). For a single random variable
this algebra is isomorphic to the one--sided shift; for a finite
stochastic process (a random variable with values in a finite
dimensional vector space) it is isomorphic to an extension, by the compact
operators,
of the {\it Cuntz algebra\/} $O_n$ (the $C^*$--algebra generated by
$n$ isometries with orthogonal ranges); for an infinite stochastic process
($card (T) = \infty$) the Toeplitz and the Cuntz algebra coincide.

The definition of random variable given above shall be sufficient for our
purposes, but it should be remarked that, it only includes those random
variables which have finite moments of all orders.  Thus, for example,
the Cauchy (Lorentz) distribution, ubiquitous in exponential decays, or
the stable laws, which receive more and more attention in the study of
chaotic systems, are excluded from this approach.

If one wants to
include all random variables one needs the following more subtle
defintion (cf. [AFL82]).
An {\bf algebraic random variable} is a triple
$$ \{ \{ {\cal A},\ \varphi\}, \ {\cal B},\ j\}\eqno(2.4)$$
where $\{{\cal A},\ \varphi\}$ is an algebraic probability space
({\bf sample algebra}); ${\cal B}$ is a $^*$--algebra ({\bf state algebra})
and $j:{\cal B}\rightarrow {\cal A}$ is an embedding. In the following,
unless explicitly stated, we shall assume that $j$ is a *-homomorphism,
i.e. $j(1_{\cal B}) = 1_{\cal A} $. When this is not the case, we shall
speak of a {\bf non
conservative } algebraic random variable. If ${\cal B}$ is
commutative we speak of a {\bf classical } algebraic random variable. If
${\cal B}$ is non-commutative, of a {\bf quantum random variable}.
The state $\varphi_j$, on ${\cal B}$, is defined by:
$$\varphi_j:=\varphi\circ j \eqno(2.5)$$
is called the {\bf distribution} of the random variable $j$. As usual a
stochastic process is a family of random variables $(j_t)$ indexed by a
set $T$. The process is called {\it classical } if ${\cal B}$ is
commutative and the algebras $j_t({\cal B})$ commute for different $t$.
If $x_j$ ($j\in F$) is a set of generators of the algebra ${\cal B}$,
we define the new index set $T' := F\times T $ and the set
$\{ X_{(j,t)}=j_t(x_j)
\ : \ (j, t)\in T' = F \times T \} $ is a stochastic
process in the sense of the previous definition.

The finite dimensional joint correlations of a stochastic process are
the quantities
$$ \varphi (j_{t_1}(b_1) \ldots j_{t_n}(b_n))\eqno(2.6)$$
for all $n\in {\bf N} $, $b_1\ldots ,b_n \in {\cal B}$ and
$t_1\ldots,t_n \in T$. It can be shown that they uniquely
characterize the stochastic process up to isomorphism (cf. [AFL82]):
this a generalization of the Kolmogorov reconstruction theorem of a
stochastic process in terms of the finite dimensional joint
probabilities as well as of the Wightman reconstruction theorem.

Moreover one can prove that the convergence of the correlation kernels
reduces, when one fixes a set of generators, to the notion of
convergence in distribution, as introduced above.\bigskip

\noindent{\it Remark\/}. The choice of a set of generators corresponds,
in probability, to the use of local coordinates in geometry: it is
useful to do calculations, but it is not intrinsic. The more general,
purely algebraic, point of view, described at the end of the present
section, provides a more intrinsic invariant approach to
stochastic processes.\bigskip

\beginsection{(3.) Statistical Independence}

Mathematically the notion of {\it statistical independence\/} is related
to that of product: classical (Boson) independence -- to the usual tensor
product; Fermion independence -- to the ${\bf Z}_2$--graded tensor product;
quantum group independence -- to more general forms of twisted tensor
products (cf. [Sch\"u93]; free independence -- to the free product.

{\it Statistical independence\/} means, in physical terms, {\it absence
of interaction\/}. Notice that also in absense of interaction one can
have {\it kinematical relations\/}, which are reflected by algebraic
constraints. For example two Fermion operators $F(f_1)$, $F(f_2)$
corresponding to orthogonal test functions, are independent in the
vacuum state, but related by the CAR: $F(f_1)F(f_2)=-F(f_2)F(f_1)$.

In classical probability one can {\it encode\/} all the algebraic
relations into the state (thus making than statistical relations). The
analogous result for quantum probability is the {\bf reconstruction
theorem} of [AFL82]. Concrete examples of such encodings are the deduction
of the CCR (resp. the CAR from a boson (resp. Fermion) gaussian state
[GivWa78], [vWa78]   (cf. Section (3.) below) deduction of the Cuntz algebra
from a free--gaussian state [Fag93].

In view of this {\it we conjecture that the natural generalizations of
the Heisenberg commutation relations shall be deduced from the GNS
representation of stochastic processes in the sense of quantum
probability\/} (independent processes, as generalizations of free
systems, Markov processes as generalizations of interacting systems).
More precisely {\it the various possible notions of free physical
system\/}. For this reason an investigation of the notion of statistical
independence is relevant for physics.

The following definition includes all the notions of independence
considered up to now:

\bigskip
\noindent{\bf Definition  ( )}.
Let $\left\{{\cal A}, \varphi\right\}$ be an algebraic probability space;
$T$ a set; $\left({\cal A}_t\right)_{t\in T}$ a family of subalgebras of
${\cal A}$; ${\cal F}$ a subset of
$ \bigcup_{n=0}^\infty T^n $
i.e. ${\cal F}$ is a family of ordered $n$--tuples of elements of $T$;
and, for each $(t_1,\dots,t_n)\in{\cal F}$, let be given a family of
subsets of the set $\{t_1,\dots,t_n\}$.

The algebras $\left({\cal A}_t\right)_{t\in T}$ are called
$({\cal F} \ , \varphi )$-{\bf independent} if
for any $(t_1,\dots, t_n )\in  {\cal F} $ and  for any choice of
$a_{t_j} \in {\cal A}_{t_j}(X)\ \  (j=1,\dots, n)$ such that, for some
$\{t'_1,\dots,t'_m\}\in{\cal F}$ one has:
$$\varphi\left(a_{t'_j} \right) = 0\eqno(3.0)$$
for all $j=1,\dots,m$, then one also has
$$\varphi\left(a_{t_1}\cdot\dots\cdot a_{t_n}\right) = 0\eqno(3.1)$$
It is moreover required that
for each $n$--tuple $(t_1,\dots,t_n)$, the family $F(t_1,\dots,t_n)$
includes all the {\it singletons\/} of the set $\{t_1,\dots,t_n\}$
(i.e. those subsets consisting of a single element $t_1\in\{t_1,\dots,t_n\}$.
The classical, Boson and Fermion independence correspond to the case in
which:
\item{i)} ${\cal F}$ is the set of $n$--tuples
$(t_1,\dots,t_n)\in T^n(n\in{\bf N})$ such that $t_i\not=t_j$ for all
$i\not=j \quad (i,j=1,\dots,n)$
\item{ii)} for each $n$--tuple $(t_1,\dots,t_n)$ the family
$F(t_1,\dots,t_n)$ consists of all the {\it singletons\/} of the set
$\{t_1,\dots,t_n\}$.

If in a general product $a_{t_1}\cdot\dots\cdot a_{t_n} $ one adds and
subtracts from $a_{t_1}$
its $\varphi $--expectation, condition (3.1) implies, that
$$\varphi\left(a_{t_1}\cdot\dots\cdot a_{t_n}\right) =
\varphi\left(a_{t_2}\right)
\cdot\varphi\left(a_{t_2}\cdot\dots\cdot a_{t_i}\cdot
\dots\cdot a_{t_n}\right)\eqno(3.2)$$
Notice that, as far as indepndence is concerned, there is no difference
between the three cases. However, in the classical case the algebras
${\cal A}_t$ are commutative and commute for different $t$; in the Boson
case they are noncommutative but still commute for different $t$; in the
Fermi case they are noncommutative and anticommute for different $t$.
\bigskip
The {\it free independence\/}, introduced by Voiculescu, corresponds to
the case in which:
\item{i)} ${\cal F}$ is the set of $n$--tuples $(t_1,\dots,t_n)\in T^n$
$(n\in{\bf N})$ such that $t_i\not=t_{i+1}$, $i=1,2,\dots,n$.
\item{ii)} The subset $F$ of $\{t_1,\dots,t_n\}$ coincides with
$\{t_1,\dots,t_n\}$ itself (this means that, for the expectation (3.1)
to be zero, all the $a_{t_i}$ must have mean zero).

Notice that the factorization property of the classical independent random
variables:
$$\varphi(a_{t_1} \cdot\dots\cdot
a_{t_n})=\varphi(a_{t_1})\cdot\dots\cdot\varphi(a_{t_n})$$
whenever $\varphi(a_{t_j})=0 $ and $t_j \ne  t_k $ for $j\ne  k $,
follows by induction for all the notions of independence discussed here.

Intermediate cases have not yet been studied and in fact it is not
clear, up to now, if other cases are possible (cf. [Sch\"u94] for a
general analysis of the motion of independence); for other
generalizations of the notion of independence cf  [BoSp91], [SpevW92].
\bigskip

Notice that, in general, the notion of
independence alone is not sufficient to
determine the form of all the correlation functons, i.e. the equivalence
class of the process. In fact it gives no information about the
expectations of products of the form $a_{t_1}\cdot\dots\cdot a_{t_n} $
with $(t_1,\dots,t_n ) \notin  {\cal F} $.

In the usual cases however this reconstruction is possible due to
commutation relations or to stronger conditions (as in the free case).

Comparing the formulation of condition (i) in the usual independence and
in the free one, one sees that the notion of free independence is {\it
essentially non commutative\/}.

For example the condition $\varphi(a_sa_ta_sa_t)=0$ if
$\varphi(a_s)=\varphi(a_t)=0$ and $s\not=t$, would imply, if the
algebras ${\cal A}_s$ and ${\cal A}_t$ commute (or anticommute), that
$\varphi(a^2_sa^2_t)=0$ which implies $a_s=a_t=0$, if both $a_s$ and
$a_t$ are self--adjoint and the state $\varphi$ faithful (strictly
positive on positive non zero elements).

Given a stochastic process $(X_t)_{(t\in T)}$ in a probability space
$\{{\cal A},\varphi\}$, one can associate to each random variable $X_t$
the $*$--algebra ${\cal A}_t$, generated by $X_t$ and the identity.

The random variables $(X_t)(t\in T)$ are called free--independent (or
simply {\t free\/}) if the corresponding algebras ${\cal A}_t$ are free
independent in the sense of the above definition.\bigskip

\beginsection{(4.) Free Fock space and $q$--deformations}

\bigskip
% {\it On the diagrams\/}:\bigskip

Let us now consider the Boson $(B)$, Fermi $(F)$ and free $(Fr)$ vacuum
expectations
$$\langle\Phi,A^{\varepsilon(1)}(f_1)A^{\varepsilon(2)}(f_2)\dots
A^{\varepsilon(N)}(f_N)\Phi\rangle_V\eqno(4.1)$$
where $N\in{\bf N}$, $\varepsilon\in\{0,1\}^N$,
$$A^\varepsilon=\cases{
A\quad&$\hbox{if }\varepsilon=0$\cr
A^+\quad&$\hbox{if }\varepsilon=1$\cr}$$
$V\in\{B,F,Fr\}$, and $\langle,\rangle_B$ (resp. $\langle,\rangle_F$,
$\langle,\rangle_{Fr}$) denotes the Boson (resp. Fermi, Free)--vacuum
expectation on the Boson (resp. Fermi, Free) Fock space over a certain
Hilbert space $H$.

A common feature of the three cases in that (4.1) is equal zero if
$N=2n+1$ is odd.

In the following we shall investigate (4.1) for $N=2n$. Another common
feature of the three cases is that (4.1) is equal to zero if either
$\exists\,r\leq2n$, such that, denoting $|\cdot|$ the cardinality of a
set
$$|\{k:k\leq r,\ \varepsilon(k)=0\}|<|\{k:k\leq r,\
\varepsilon(k)=1\}|\eqno(4.2a)$$
or
$$|\varepsilon (k) :k\leq2n,\ \varepsilon(k)=0\}|\not=|\{\varepsilon
(k):k\leq 2n,\ \varepsilon(k)=1\}|  \eqno(4.2b)$$
For each $\varepsilon\in\{0,1\}^{2n}$ such that (4.2a) and (4.2b)
are not true.
We connect with a line a vertex  $k\in\{i:\varepsilon(i)=0\}$ and a
vertex $m\in\{i:\varepsilon(i)=1\}$. Thus we need $n$--lines to connect
the $2n$ vertices and in the following we call this a {\it connected
diagram\/}.

Let us denote by $1=p_1<p_2<\dots<p_{2n-1}<p_{2n}<2n$, the vertices such
that $\varepsilon(p_k)=0$, $k=1,2,\dots,n$ (i.e. the creator vertices).
Then by the CCR or the CAR, (.1) is equal to
$$\sum_{1<q_1,\dots,q_n=2n\atop{q_h>p_h,h=1,\dots,n\atop{\{q_h\}^n_{h=1}=
\{1,\dots,2n\}\backslash\{p_h\}^n_{h=1}}}}\kappa_\varepsilon\prod^n_{h=1}
\langle f_{p_h},f_{q_h}\rangle\eqno(4.3)$$
where, for $V\in\{F,Fr\}$, $\kappa_\varepsilon\in\{0, \pm 1\}$ is determined
uniquely by $\{p_h,q_h\}$ and it is always equal to $1$ in the case $V=B$.

In (4.3), for each choice of $\{q_h\}^n_{h=1}$ we have a diagram with
lines $\{(p_h,q_h)\}^n_{h=1}$. In other words, we have a pair--partition
$\{(p_h,q_h)\}^n_{h=1}$ of $\{1,2,\dots,2n\}$. For fixed
$\{p_n\}^n_{h=1}$, in the cases $B$ and $F$, we have many choices of
$\{q_h\}^n_{h=1}$'s and each choice determines a diagram.
But for each fixed $\{p_h\}^n_{h=1}$ there is at most one choice of
the $\{q_h\}^n_{h=1}$ such that the associated
diagram is non--crossing, i.e. for all   $1\leq k\not=h\leq n$,
$$p_k<p_h\Rightarrow q_k<p_h\eqno(4.4)$$
Thus, for $V=Fr$, (4.1) differs from zero only if $\varepsilon$
allows a non--crossing
diagram $~~~~~~~$ $\{(p_h,q_h)\}^n_{h=1}$ and if so (4.1)
is given by
$$\prod^n_{h=1}\langle f_{p_h},f_{q_h}\rangle \eqno(4.5)$$
In this case the distribution of the field operators $ A(f)+A^+(f)$,
$f\in H $ is a joint semi--circle distribution (in the sense that the
marginals are semi--circle laws with variance proportional to the square
norm of the test functions). In Section (12.) we shall see that the
stochastic limit of quantum electrodynamics sugggests a nonlinear
modification of the expression (4.5) which no longer gives the
semi--circle law for the field operators, but still keeps the structure
of the non--crossing diagrams.

\bigskip
The $q$--deformed commutation relations in the full Fock space
${\cal F}(H)$ (cf. for example [APVV]
and references therein
$$c(f)^*(g)-qc^*(g)c(f)=\langle f,g\rangle$$
interpolate between Bose relations ($q=1$), Fermi ($q=-1$) and free
($q=0$).

For $-1<q<1$ one gets the relation [BoSp91]
$$\langle\Phi,(c(f)+c^*(f))^{2n+1}\Phi>=0$$
$$\langle\Phi,(c(f)+c^*(f))^{2n}\Phi=\int_{2/\sqrt{1-q}}x^{2n}\nu_q(x)dx$$
where
$$\eqalign{
\nu_q(x)&={\sqrt{1-q}\over\pi}\,\sin\hbox{ Arccos }
{\sqrt{1-q}\over2}\,\cdot\prod^\infty_{k=1}(1-q^k)\cdot\cr
&\cdot\left|1-q^k\exp\left\{-2i\hbox{ Arccos
}{\sqrt{1-q}\over2}\,x\right\}\right|^2\cr}$$
In particular for $q=0$ one has the Wigner semicircle distribution,
$\nu_0=w_{1/2}$ and
$$\nu_q(x)\to{1\over\sqrt{2\pi}}\,\exp\left(-{x^2\over2}\right),\
q\to1(\hbox{Bose})$$
$$\nu_q(x)\to{1\over2}\,(\delta(x-1)+\delta(x+1)),\ q\to-1\hbox{
(Fermi)}$$
For the free algebra (4.1) one has
$$\langle\Phi,(a(f)+a^*(f))^{2n+1}\Phi\rangle=0$$
$$\langle\Phi,(a(f)+a^*(f))^{2n}\Phi\rangle=x_{2n}\cdot\|f\|^{2n}$$
where $x_{2n}$ is the number of non--crossing partitions of
$\{1,\dots,2n\}$, the so--called Catalan number,
$$x_{2n}\cdot\|f\|^{2n}=\int_{{\bf R}^N}x^{2n}W_{\|f\|}(x)d
x$$
\bigskip

\beginsection{(5.) Gaussianity (classical, Boson, Fermion, Free, $\ldots$) }

\bigskip

The results of the central limit theorems suggest the independent study
of the class of maps which canonically arise from these theorems, these
are the so--called {\it Gaussian maps}.

Let ${\cal B}$ and ${\cal C}$ be *--algebras
and let $E:{\cal B}\longrightarrow {\cal C}$ be a linear $*$--map. A
family $B\subseteq {\cal B}$ is called a {\bf mean zero generalized Gaussian
family } with respect to $E$, if for each $n\in{\bf N}$ and for each sequence
$b_1,\,\dots\,, b_n$ of elements of $B$, not necessarily different
among themselves, one has:
$$E(b_1\cdot\dots\cdot b_n)=0\; \ {\rm if}\; n\;{\rm is\; odd}\eqno(5.1)$$
and for each even $n=2p$, there exists a subset ${\cal P}{\cal P}_o(n)$,
of the set ${\cal P}{\cal P}(n)$ of all the ordered pair partitions of the set
$\{ 1, \ldots , n =2p \} $ such that
$$E(b_1\cdot\dots\cdot b_{2p})={1\over p!}
\sum_{(i_1,j_1,\ldots ,i_p,j_p) \in {\cal P}{\cal P}_o(n)  }
\epsilon (i_1,j_1;\,i_2,j_2;\,\dots\,; i_p,j_p)E(b_{i_1} b_{j_1})
\cdot\dots\cdot  E(b_{i_p}\cdot b_{j_p})\eqno(5.2)$$
where the pair partition $(i_1,j_1;\,i_2,j_2;\,\dots\,; i_p,j_p)$ is called
ordered if a
permutation of the pairs $(i_\alpha,\,j_\alpha)$ changes the partition while
$$i_ \alpha <j_ \alpha \qquad ; \qquad \alpha =1,\dots, p $$
and for each natural integer $n$, $\epsilon _n $ is a complex valued character
of the permutation group over $2n$ elements, e.g.
$\epsilon _n (j_1,\dots,j_{2n}) = + 1 $ for all permutation (Boson case)
or $\epsilon _n (j_1,\dots,j_{2n}) = sgn (j_1,\dots,j_{2n})  $
(Fermion case).

Notice that positivity is not required a priori, but the most intresting
examples are obtained in correspondence of positive (even completely
positive) maps.
It is not known for which subsets ${\cal P}{\cal P}_o(n)$ of pair partitions
these prescriptions define a positive functional. The standard Gaussian
(Boson or Fermion) case corresponds to the choice of all pair
partitions; the free case corresponds to the so--called {\it non
crossing} (or rainbow) partitions, which are characterized by the
condition:
$$ i_k < i_h < j_k\leftrightarrow i_k < i_h < j_h  < j_k\eqno(5.3)$$
in other terms: {\it the intervals corresponding to two pairs }
$(i_k, j_h ) $ and $(i_h , j_k) $ {\it are ither disjoint or one is
contained in the other\/}.

Other subsets which give positivity can be constructed by hands, but it
is not known if they come from some central limit theorem.

One easily checks that generalized gaussianity is preserved under linear
combinations in the sense that, if $B\subseteq{\cal B}$ is an $E$--Gaussian
family then the linear span $[B]$ of $B$ is also an $E$--Gaussian family.
The sesquilinear map on $[B]$
$$q(b_1,b_2) := E(b^*_1 \cdot b_2) $$
is called the {\bf covariance} of the Gaussian map $E$.

If ${\cal C} = {\bf C}$ = the complex numbers,
$E$ is positive and $E(1) =1$ , then $E$ is called a {\bf Gaussian state}.
In the physical literature, Boson and Fermion Gaussian states are
called {\bf quasi-free} states.

In case of Boson Gaussian states (which include the classical
probability measures) one easily checks the well known identity for Gaussian
measures i.e., in case of a single self--adjoint random variable $b$:
$$ E(e^{zb} ) = e^{ {1\over 2} z^2 E(b^2) } \eqno(5.4)$$
In case of a single random variable, if ${\cal P}{\cal P}_0(n)$ is the
set of non crossing partitions and $\varepsilon$ is identically equal to
1, then condition (5.12) gives the momenta of
the {\bf Wigner semicircle law}.
\bigskip
The following theorem, essentially due to von Waldenfels,
shows that the canonical commutation and anticommutation relations have
their deep root in the notion of Gaussianity.
\bigskip
\noindent{\bf Theorem 5.1}.
{\sl Let $ \{{\cal A },\varphi \}$
be an algebraic probability state.
Let $B$ be a family of algebraic generators of $ {\cal A } $ and suppose that
$B$ is a mean zero (Boson or Fermion) Gaussian family with respect to
$\varphi    $ in the sense of the above definition}.\bigskip

Denote, for $a,b \in {\cal A} $ , for any natural integer $n$ and
$j_1,\dots,j_n \le n $ natural integers :
$$[a,b]_\epsilon =
\cases{  ab-ba , & if\ $\varphi $ is Boson-Gaussian \cr
   ab+ba , & if\ $\varphi $ is Fermion-Gaussian \cr}
\eqno(5.5)$$
and let $\{ {\cal H} , \pi , \Phi \}$ be the GNS representation of the
pair $ \{ {\cal A} , \varphi \} $.
Then for any pair $b_1, b_2 \in B$ one has
$$[\pi  (b_1), \pi (b_2)]_\epsilon = \varphi    ( [ b_1 , b_2 ]_\epsilon  )
\eqno(5.6)$$
\bigskip
The above theorem suggests to look at the GNS representation of Gaussian
(or even more general) states as a natural source of generalized
commutaion relations. In other terms: {\it up to now the commutation or
anticommutation relations have been related to the kinematics (e.g group
representations)\/}. The {\bf locality condition} itself should find its
natural formulation as an expression of some form of {\it statistical
independence\/} rather than in terms of commutation relations: the
former can be formulated uniquely in terms of observable quantities; the
latter is strongly model dependent.
{\it The above theorem suggests that their deeper root is
statistic\/}. The statistical root is
related to a universal phenomenon such as the central limit theorem and
the theory of unitary group representations should enter the picture as
a theory of {\it statistical symmetries\/}.

In particular one might ask oneself which are the commutation relations
coming from free gaussianity. This problem has been solved by Fagnola
[Fag93]  who proved that essentially they are the Cuntz relations.
More precisely, in full analogy with the Boson and the Fermion case,
in the case of a single non self--adjoint random variable $a, a^+$,
if the rank of the covariance matrix is one, the GNS representation is
unitarily equivalent to the free (full) Fock representation described in
Section (4.); if it is two, then it is unitarily isomorphic to the free
product of a free Fock and a free anti--Fock representation. The analogy
with the Boson and the Fermion case suggests to interpret this latter
type of states as {\it the free analogue of the finite temperature
states}. A strong confirmation to this conjecture would come if one could
prove that, in the stochastic limit of QED without dipole approximation,
if the field is taken in a finite temperature representation, rather
than in the Fock one, as in [AcLu92],   one obtains not the interacting
Fock module but the corresponding finite temperature module
(cf. Section (12.) below for the terminology).\bigskip

\beginsection{(6.) Algebraic laws of large numbers and central limit theorems}

\bigskip
Let $ \{ {\cal A}, \varphi \}$ be an algebraic probability space and let
$j_n:{\cal B}\rightarrow {\cal A}$ ($n\in {\bf N}$) be an algebraic stochastic
process. The laws of large numbers (LLN) study the behaviour of the sums
$$ {1 \over N } S_N(b):={1 \over N } \sum^N_{n=1}j_n(b)\eqno(6.1)$$
for some $b\in {\cal B} $, and the central limit theorems (CLT), the behaviour
of the centered sums
$$ {1 \over \sqrt N } \bar S_N(b):={1 \over \sqrt N } \sum^N_{n=1}
[j_n(b)  - \varphi (j_n(b)) ]\eqno(6.2)$$
In case of a vector or matrix valued ($k$--dimensional) random variables, one
studies the sums (6.1), (6.2) not for a single $b\in {\cal B} $, but for
a $k$--tuple of elements
$$ b^{( 1)}, \ldots , b^{( k)} \in {\cal B} $$
Thus introducing, for $q=1,\dots,k$, the notation
$$b^{( q)}_n :=\cases{
j_n(b^{(q)})\qquad\hbox{LLN--case}\cr
[j_n(b(q))  - \varphi (j_n(b^{(q)}))]\quad
\hbox{CLT--case}\cr}\eqno(6.3)$$
one can reduce both the laws of large numbers and the central limit theorems,
to the study the behaviour of the sums
$$ {1 \over N^\alpha  } S_N(b^{( q)}):=
{1 \over N^\alpha  } \sum^N_{n=1}(b_n^{( q)}) \qquad ; \ q=1,
\ldots , k\eqno(6.4)$$
More precisely, one is interested
in the limit, as $N \rightarrow \infty$ of expressions of the form
$$
\varphi \left ( P \left [ {S_N(b^{( 1)}) \over N^{\alpha}} ,
{S_N(b^{(2)}) \over N^{\alpha}} ,
\ldots , {S_n(b^{( k)}) \over N^{\alpha}} \right ] \right ) \eqno(6.5)$$
where $\alpha > 0$ is a scalar, $b^{( 1)},\ldots ,b^{( k)}\in B$ and $P$ is a
polynomial in the $k$ noncommuting indeterminates $X_1,\ldots ,X_k$.
The law of large numbers is obtained when $\alpha = 1$
and the central limit theorem when $\alpha =1/2$.

This convergence is called {\bf convergence in the sense of moments}.
In the classical (commutative) case, one replaces the polynomial $P$ in
(6.5) by a continuous bounded function, and obtains the (strictly stronger)
notion of {\bf convergence in law}; however if the $b^{( q)}_n)$ do not
commute there is no natural meaning for the expression (6.5) if $P$ is not a
polynomial (or an holomorphic function, if a topology is given on ${\cal A}$).

By linearity one can replace the polynomial $P$ in (6.5) by a
noncommutative monomial and the study of expressions of the type ( ) is
reduced to the study of expressions of the type
$$
\varphi \left ( {S_n(b^{( 1)}) \over N^{\alpha}} \cdot
{S_n(b^{( 2)}) \over N^{\alpha}} \cdot
\cdots \cdot {S_n(b^{( k)}) \over N^{\alpha}} \right ) \eqno(6.6)$$
where $b_1,\ldots ,b_k \in {\cal B}$ and it is not required that
$b^{( i)}_n \ne b^{( h)}_n$ if $h \ne i$.

The stochastic process $j_n : {\cal B} \rightarrow {\cal A}$
is called {\it homogeneous } (or stationary) if
$$ \varphi \circ j_k = \varphi \circ j_o =:
\varphi_o  \qquad ; \quad \forall k \in {\bf N}  \eqno(6.7)$$
and this notion can be extended to the case when $\varphi$ is a general
linear map
from $ {\cal A}$ to a subalgebra ${\cal C}$ rather than a state on ${\cal A}$.
In fact almost all the results of this Section extend to maps

As in classical probability, also in algebraic probability the laws of
large numbers and the central limit theorem are proved under
independence assumptions. Here however one has to distinguish between
the various types of independence which are possible.
\bigskip

In the above notations, a linear map $\varphi: {\cal A} \to {\cal C}$
is called {\bf factorizable } or simply a {\bf product map}, if
for each integer $n$ and $b_1,\dots,b_n \in { \cal  B } $  one has
$$\varphi(j_1  (b _1)\cdot \dots \cdot j_n (b_n ) )
 = \varphi(j_1  (b _1) ) \cdot \dots \cdot \varphi(j_n (b_n))\eqno(6.8)$$
Notice that factorization is required only in the case when the indices are
in {\bf strictly increasing} order. If moreover the stationarity condition
( ) is satisfied
then $\varphi $ is called a {\bf homogeneous product map} with
marginal $\varphi_o$.
\bigskip

\noindent{\bf Theorem 6.1}.
{\sl Let ${\cal C} \subseteq \hbox{Center}({\cal A})$ and
$\varphi:{\cal A}\rightarrow {\cal C} $ be a
${\cal C}$-linear product map and suppose that,
for $j\ne k$, $j_h({\cal B}) $ and $j_k({\cal B}) $ commute (Boson case)
and that ${\cal A} $ is generated by
$$\{{\cal C} \lor j_k({\cal B})\; :\; k=1,2,\dots\}$$
Then, if the limit:
$$\lim_{N\rightarrow\infty}\varphi\left({S_N(b)\over N}\right)=
\lim_{N\rightarrow\infty}{1\over N}\sum^N_{k=1}\varphi(j_k(b))=:
\varphi_o(b)\eqno(6.9)$$
exists in the norm topology of ${\cal C}$ for each $b\in B $, it follows
that for any natural integer $k$, for any $b_1,\,\dots\,, b_k\in {\cal B}$,
and for any polynomial $P$ in $k$ non commuting indeterminates,
one has
$$\lim_{N\rightarrow\infty}\varphi\left( P\left( {S_N(b_1)\over N}, \dots,
{S_N(b_k)\over N}\right)\right)=
P\left(\varphi_o(b_1),\,\dots\,, \varphi_o(b_k)\right)\eqno(6.10)$$}
\bigskip

The following is the simplest algebraic central limit theorem. It extends
the classical central limit theorem for independent identically distributed
random variables to the Boson independent case.
\bigskip
\noindent{\bf Theorem 6.2}. {\sl Let
$\varphi_o : { \cal  B} \longrightarrow { \cal  C} $ be a map and let
$\varphi $ be an homogeneous product map on ${ \cal  A}$ with
marginal $\varphi_o$ . Suppose moreover that,
for $j\ne k$, $j_h({\cal B}) $ and $j_k({\cal B}) $ commute
Then,  for any $k\in \bf N $ and any elements
$b_1 ,\dots., b_k \in {\cal B}$ such that
$$\varphi_o (b_j  ) = 0  \qquad ; \qquad j=1,\dots,k \eqno(6.11)$$
one has
$$\lim_{N\to \infty }\varphi\Biggl({  S_N(b_1) \over N^{1/2} } \cdot { S_N(b_2)
\over N^{1/2  } }  \cdot \dots \cdot
{ S_N(b_k) \over N^{1/2} } \Biggr) = 0 \eqno(6.12)$$
if $k$ is odd  and, if $k = 2p $ for some integer $p$ then
$$\lim_{ N \to \infty }
\varphi\Biggl({  S_N(b_1) \over \sqrt N } \cdot { S_N(b_2)
\over \sqrt N }  \cdot \dots \cdot { S_N(b_k) \over \sqrt N  }
\Biggr) = $$
$$= \sum_ { (S_1,\dots,S_p) \in { \cal  P } _{ p,2p} }
{ 1 \over p! } \sum _ { \pi \in { \cal  S } _p }
\sigma ( \pi ,S_1,\dots,S_p ,b_1,\ldots ,b_k)
\varphi_o(b_{  S_{ \pi (1) } } )\cdot \dots \cdot
\varphi_o(b_{  S_{ \pi (p) } } )\eqno(6.13)$$
where $\sigma (\pi ,S_1,\dots,S_p ,b_1,\ldots ,b_k) $ is a constant depending
only on the permutation $\pi $, on the partition $ ( S_1,\dots,S_p )$ of
$(1, \ldots ,k)$ and on $ b_1,\ldots,b_k$}.

\bigskip
\noindent{\it Remark\/}.
A partition $(S_1,\dots,S_p ) $ of the set $\{1,\dots,k \}$ will be
said to  contain a {\bf singleton} if,  for  some  j=1,\dots..,p,
$S_j$  consists of a single element.
The basic idea in the proof of the law of large numbers and of
the central limit theorems for product maps, with the method of moments,
is that the partitions are divided into two classes:
those with a singleton and those without. Those with a singleton
give zero contribution because of (10.1) and the product
map assumption. For a partition $(S_1,\dots ,S_p)$, with no singleton,
$p$ must be $\leq k/2$ and the number of terms in the
sum is of order $N^p$, hence it is balanced by the normalizing
factor $N^{\alpha k}$. Thus, in the mean zero case, a finite nontrivial
limit can only take place if $\alpha=1/2$ and in this case $p$ must be
equal to $k/2$, i.e. $(S_1,\dots ,S_p)$ must be a pair partition.
\bigskip
The {\it invariance principles} (or functional central limit theorems)
study the sums\hfill\break
${  S_{ [Nt] } (b) / N^{1/2}}$, where $t$ is a
real number and $[Nt]$ denotes the integer part of $t$, and show that,
under natural conditions they converge to some (classical or quantum)
Brownian motion. Recalling that
$$ {1  \over N^{1/2} } S_{ [Nt] } (b) =
{1\over N^{1/2} } \sum_{k=1}^{ [Nt] } j_k(b)
={1\over N^{1/2} } \sum_{k=1}^{ [Nt] } \int_{k-1}^{ k } j_k(b) ds = $$
$$ ={1\over N^{1/2} } \sum_{k=1}^{ [Nt] } \int_{k-1}^{ k }
\chi_{[k-1,k]}(s) j_k(b) ds$$
where $\chi_{[k-1,k]}(s) = 0$ if $s\in [k-1,k]$ and $=1$ if $s\notin [k-1,k]$,
one can generalize the above sums by considering limits, as $ \lambda \to 0 $
of expressions of the form
$$ \lambda \int_{S^2/\lambda^2 }^{T^2/\lambda^2 }X(s)
ds=S_\lambda([S,T],X)$$
where $X(s) $ is an operator valued function. These are precisely the
kind of limits that one meets in the stochastic limit of quantum field
theory (cf. Section (10.) below).

In their full generality, the first Boson and Fermion central limit
theorems were proved by von Waldenfels [GivWa78], [vWa78] (cf. also
pioneering work by Hudson [Hu71], [Hu73]. Voiculescu [Voi91] proved the first
free central limit theorem. Speicher [Sp90]  showed that the same techniques
used in the proof of the usual quantum central limit theorems allow to
prove also the free central limit theorem.  The first quantum invariance
principle was proved by Accardi and Bach [AcBa87]  in the Boson case and by
Lu [Lu89]  in the Fermion case.\bigskip

As already said, independence corresponds to free systems, whose
physics is usually not very interesting. Interaction corresponds to
statistical dependence. However, as in classical statistical mechanics
an Euclidean field theory, the locality properties of the interactions
allow to express nontrivial informations in terms of local perturbations
of the free fields. In probabilistic language, this leads to the notion
of {\it weak dependence\/} (manifested by a decay of correlations).
Under weak dependence assumptions, the CLT still holds (cf. [Verb89]  for
the CCR case and [AcLu90b]  for general $q$--deformed commutation
relations.\bigskip\bigskip

\noindent{\bf (7.) Large $N$ limits of matrix models and asymptotic freeness}

\bigskip
We try to explain here why it is natural to expect, as suggested by
Singer [Sin94], that
some results of Voiculescu on random matrices might help clarifying
the $1/N$ expansion of matrix models and its extensions.

The basic idea can be summarized in the following statement: a set
$A^{(N)}_1,\dots,A^{(N)}_k$ ($k$ can be infinite) of $N\times
N$ symmetric random matrices with independent (modulo symmetry) gaussian
entries, with mean zero and variances of order $1/N$, {\bf tend to
become free random variables as $N$ becomes large} $(N\to\infty)$.

The physical interpretation of this phenomenon is best understood in
terms of chaos: {\bf freenes is an indication of chaotic behaviour}. In
fact, just as statistical independence means {\it absence of statistical
relations\/}, so freeness means {\it absence of algebraic relations\/}.
Therefore, in some sense, {\it free independence\/} denotes a maximum of
chaoticity.

In the $1/N$--expansion one is precisely concerned with the limit, as
$N\to\infty$, of $N\times N$ matrix models. Truly,  the large matrices
are not always hermitean (e.g. they can be unitary) and their
distributions are not indpendent gaussian because of the interaction
and of the algebraic constraints.

However intuitively one expects that if the deviation from gaussianity
is not too large (cf. Theorem 5.2 below) the general picture should not
change too--much.

In the following we review some definitions and theorems that make this
picutre more precise.

Let $\{{\cal A}_N,\varphi_N\}$ be a sequence of algebraic
probability spaces and, for each $N$ let $A^{(N)}=(A^{(N)}_i)_{i\in I}$ be a
family of random variables in ${\cal A}_N$. The sequence of joint
distributions $\varphi_{A^{(N)}}$ converge as $N\to\infty$, if
there exists a (state) $\tilde\varphi$ such that
$$\varphi_{A^{(N)}}(P)\to\tilde\varphi(P)$$
as $n\to\infty$ for every $P$ in ${\bf C}\langle\tilde X_i|i\in I\rangle$
(algebra of
polynomials on noncommuting indeterminants $X_i$) $\tilde\varphi$ is
called the limit distribution.

Let $I=\bigcup_{j\in J} I_j$ be a partition of $I$.

A sequence  of stochastic processes $(\{A^{(N)}_i|i\in I_j\}_{j\in J}$
is {\it asymptotically free\/} as $N\to\infty$
if it has a
limit distribution $\tilde\varphi$ and if $(\{X_i|i\in I_j\})_{j\in
J}$ is a free--independent stochastic process in the probability space
$\{{\bf C}\langle X_i|i\in I\rangle,\tilde\varphi\}$.

Let $(\Omega,{\cal F},P)$ be a classical probability space denote, $L$ the
algebra $\bigcap_{1\leq p<\infty}L^p(\Omega)$ equipped with the expectation
functional $E$ given by the integration with respect to $P$.

Let $\{ {\cal M}_N,\varphi_N\}=\{L\otimes M_N({\bf C})$, $E\otimes\tau_N\}$
denote the noncommutative probability space of $N\times N$ random
matrices with entries in $L$. Here $\tau_N$ is the normalized matrix trace,
$\tau_N(A)={1\over N}\,trA$. We will consider a sequence of random
matrices $A^{(N)}_i=(A^{(N)}_{i,\alpha\beta})^N_{\alpha,\beta=1}\in{\cal
M}_N$ where $A^{(N)}_{i,\alpha\beta}$ denotes matrix elements,
$\alpha,\beta=1,2,\dots,N$ and $i=1,2,\dots$.\bigskip

\noindent{\bf Theorem 7.1}. {\sl [Voi92] . Let $A^{(N)}_i\in{\cal M}_N$,
$i=1,2,\dots$ be
hermitian random matrices such that $A^{(N)*}_i=A^{(N)}_i$ and
both $\hbox{Re }A^{(N)}_{i,\alpha\beta}$ and $\hbox{Im
}A^{(N)}_{i,\alpha\beta}$ are independent Gaussian random
variables such that for $i=1,2,\dots$ one has
$$E(A^{(N)}_{i,\alpha\beta})=0\ ,\quad1\leq\alpha\leq\beta\leq N$$
$$E((\hbox{Re }A^{(N)}_{i,\alpha\beta})^2)=E((\hbox{Im }
A^{(N)}_{i,\alpha\beta})^2)={1\over2N}\ ;\quad 1\leq\alpha<\beta\leq N$$
$$E((A^{(N)}_{i,\alpha\alpha})^2)={1\over N}\ ,\quad1\leq\alpha\leq N$$
Then the family $(\{A^{(N)}_1\},\ \{A^{(N)}_2\},\dots)$ of matrix
random variables is asymptotically free as $N\to\infty$ and moreover the
limit distribution of each $A^{(N)}_i$ is a semicircle law.}\bigskip

One can rewrite this theorem as follows
$$\lim_{N\to\infty}\varphi_N\bigl(A^{(N)}_{i_1}\dots
A^{(N)}_{i_k}\bigr) =\tilde\varphi(\tilde X_{i_1}\dots\tilde
X_{i_k})\eqno(7.1)$$
where $\tilde\varphi$ is a limiting state on ${\cal C}\langle\tilde
X_i:i\in{\bf N}\rangle$ and the $A_i$ are noncommutative
indeterminates.

If the family $(A^{(N)}_i)$ consist only of one sequence $A^{(N)}$ of
random matrices, in the same notations of the introduction, one can
write (7.1) as
$$\lim_{N\to\infty}\langle{1\over N}\,tr(A^{(N)m})\rangle_N=
\varphi(\tilde X^m)= \int_{\bf R}x^mw(x)dx \qquad ; \
m=1,2,\dots\eqno(7.2)$$
where $w(x)$ is the
Wigner semicircle distribution with unit variance. By the GNS representation
one gets an
operator realization in the free Fock space of the free algebra
${\bf C}\langle A^{\infty } _i\rangle$
In particular in the simplest case
of only one sequence $A^{(N)}$ of random matrices one gets the formulae
(1.1) and (1.5) from the introduction where $\Phi=\pi_\varphi(X)$.

There are extensions of Theorem 7.1 in different directions. In
particular there is an analogous result for a {\it standard family of
independent unitary $N\times N$ random matrices\/}. In this case a
limiting distribution is the Haar measure on the unite circle. There are
extensions also to other groups of matrices, to the fermionic case and
to the {\it non--Gaussian\/} case.\bigskip

\noindent{\bf Theorem 7.2.} {\sl [Voi92]  Let $A^{(N)}_i\in{\cal M}_N$,
$i=1,2,\dots$ be Hermitian $A^{(N)*}_i=A^{(N)}_i$ and
$A^{(N)}_{i,\alpha\beta}$ is an independent set of random variables such
that
$$E(A^{(N)}_{i,\alpha,\beta})=0$$
$$E(|A^{(N)}_{i,\alpha\beta}|^2)={1\over N}$$
$$\sup E(|A^{(N)}_{i,\alpha\beta}|^m)=O(N^{-m/2}),\ m\geq1\,,\
i=1,2\dots,1\leq\alpha\leq\beta\leq N$$}\bigskip

The the family $(\{A^{(N)}_1\} , \{A^{(N)}_2\},\dots )$ of sets of
random variables is asymptotically free as $N\to\infty$ and the limit
distribution of each $A^{(N)}_i$ is a semicircle law.\bigskip\bigskip

\noindent{\bf (8.) Quantum stochastic differential equations}

\bigskip
Let $X(f)$ be an operator depending on a test function $f$ (e.g. a field
or creation or annihilation or number operator). Suppose that $f$
depends on space--time (or time--momentum) variables, then if we denote
by $\chi_I$ the characteristic function of the time interval $I$, the
operator $X(\chi_I f)$ is localized (in time) in the interval $I$ and,
moreover, the map $I\mapsto X(\chi_I f)$ is an operator valued measure
on the real line {\bf R}. Under mild regularity conditions such a
measure is called a {\it semi--martingale} (or a stochastic integrator)
and integration with respect to measures of this
kind is called {\bf quantum stochastic integration}. Classical
stochastic integration is recovered when the algebraic random variables
$X(\chi_I f)$ commute for $I\subseteq {\bf R} $ and $f$ varying among
the test functions. In classical probability a random variable valued
measure on the real line is also called an {\it additive process}
(referring to the property $ X(\chi_{[r,s]} f) + X(\chi_{[s,t]} f) =
X(\chi_{[r,t]} f) $) and often in the following we shall use this
terminology. Often a time origin, say $t=0$ is fixed and one uses
the notations
$$ X(\chi_{[o,t]} f) =: X_t( f) $$
in these notations $ X(\chi_{[s,t]} f) = X_t( f) - X_s ( f) $ and the
notation on the right hand side is more frequently used in classical
probability. In the following we shall include the function $f$
in definition of $X$ and drop it from the notations.

Recently various kinds of stochastic calculi have been introduced
in quantum probability, corresponding to different choices of the operator
valued measures $I\mapsto X(\chi_I f)$, the first one of these being due
to
[HuPa84a], [HuPa84b]. The common features of these calculi are:
\item{--} One starts from a representation of some commutation relations
(not necessarily the usual $CCR$ or $CAR$) over a space of the
form $L^2(I,dt;{\cal K})$ where ${\cal K}$ is a given Hilbert
space and $I\subseteq {\bf R} $ is an interval.
\item{--} One introduces a family of operator valued measures on ${\bf R}_+$,
related to this representation, e.g. expressed in terms of creation or
annihilation or number or field operators.
\item{--} One shows that it is possible to develop a theory of stochastic
integration
with respect to these operator valued measures sufficiently rich to allow to
solve some nontrivial stochastic differential equations.

\medskip
The basic application of such a theory is the construction of unitary
wave operators (Markovian cocycles in the quantum probabilistic
terminology) as solutions of quantum stochastic differential
equations generalizing the usual Schr\"odinger equation (in interaction
representation). In order to achieve this goal it is necessary to prove
an operator form of the It\^o formula (whose content is that the product of
two stochastic integrals is a sum of stochastic integrals).

The exposition that follows is aimed at giving a quick idea of the basic
constructions of quantum stochastic calculus and by no means should be
considered a complete exposition (for an approach to stochastic integration
which unifies the classical as well as the several different quantum
theories, cf.
[AcFaQu90]  and [Fag90] ; for a more concrete approach, based on
the Fock space over $L^2(I,dt;{\cal K})$, cf. [Par93] ).

Let be given a complex separable Hilbert space $ {\cal H }$  and a
filtration $({\cal H}_{t]} )$ in $ {\cal H }$ (i.e. a family of
subspaces ${\cal H}_{t]} $ of $ {\cal H }$ such that, for $s\leq t$ one
has ${\cal H}_{s]} \subseteq {\cal H}_{t]} $: intuitively ${\cal H}_{t]} $
represents the history of the system up to time $t$.
We write
$ {\cal B} ( {\cal H} )$  to denote the
vector space of all bounded operators on $ {\cal H}$ .

Let $ {\cal D} $ be a total subset of $ {\cal H} $ (the linear
conbinations of elements of $ {\cal D} $ are dense).

When dealing with stochastic integrals one has to do with unbounded
random variables. In particular one cannot single out, in the quantum
case, some natural and easily verifiable conditions that assure that all
the stochastic integrals considered leave invariant a sufficiently large
domain of vectors. For this domain reasons even the general algebraic
context of $*$--algebras is too strong to deal with stochastic integrals
and one needs a weaker definition of random variable.

A {\it random variable } is a pair
$(F, F^+)$ of linear operators on $ {\cal H} $  with domain containing
$ {\cal D} $ and such that for all elements $\eta \ ,\ \xi $ in this domain
$$     <\eta  , F\xi   > = < F^+ \eta ,\xi  >$$
The (linear) space of random variables shall be denoted
$ {\cal L } ( {\cal D} ; {\cal H}  )$ and a random variable $(F, F^+)$ shall
be simply denoted $F$. A {\bf stochastic process} is a family
$(F_t) \ _{t\geq    0 }$ of random variables; it is called {\bf adapted }
to the filtration $({\cal H}_{t]} )$ if, for all  $t\geq    0$ one has
$ F_t {\cal H}_{t]}  \subseteq {\cal H}_{t]}  $. This is a {\it
causality condition} it intuitively expresses the fact that the random
variable $F_t$ does not depend on the future (with respect to $t$)
history of the system. A {\it step} (or elementary) process
is an adapted process of the form
$$F(t) = \sum _{k=1}^n \chi _{(t_k , t_{k+1} ]} (t) F^._{t_k} $$
with $0 \leq  t_o  < t_1  < ...< t_n < \infty $.

An {\it additive process} on $ {\cal H} $ is
a family $(X(s,t) ) \ ( 0\leq  s  < t) $  of
elements of $ {\cal L } ( {\cal D} ; {\cal H})$ such that, for each $s$
the process $t\mapsto X(s,t) $ is adapted and for all $r<  s  <t $ we have
$$     X(r,t) = X  (r,s)+X  (s,t) $$
Given an additive process $X$ and a step process $F$, one can define the
{\bf left stochastic integral }
$$ \int dX_s F_s = \sum _{k=1}^n X(t_k,t_{k+1} ) F_{t_k} \eqno(8.1)$$
and the right stochastic integral is defined similarly.
Stochastic integrals of step process are called {\it simple} stochastic
integrals.

An additive process $X$ is called
a {\it regular semimartingale } (or simply a stochastic integrator) for
the set $ {\cal D}$ if for all elements
$ \xi \in {\cal D}$ there exist two positive
funtions $g_{ \xi }^\epsilon \in L^1_{loc}( {\bf R}_+ ) $ ($\epsilon = 0,1$)
such that, for any step process $F$ and all $t\geq   0  $ we have:
$$ \parallel   \int _0 ^t dX_s F^\epsilon_s \xi  \parallel   ^2 \leq c_{t, \xi
} \cdot
\int _0 ^t \parallel   F^\epsilon_s \xi
\parallel   ^2 g^\epsilon _ { \xi }(s) ds \eqno(8.2)$$
where $F^\epsilon$ stands for $F$ or $F^+$ (to include free stochastic
integration one should replace $F^\epsilon$ on the right hand side of
(8.2) by $ {\cal T}_X(F^\epsilon) $ where $ {\cal T}_X $ is a linear
map depending on the integrator $X$ (cf. [Fag92] ).

The inequalities (8.2) depend on the domain on which the basic
integrators are considered. The known examples of basic integrators are
creation, annihilation and number processes with respect to a given
quasi-free (gaussian) representation of the $CCR$ or of the $CAR$.
In the Boson case they verify these inequalities on the domain of coherent
vectors but for more general domains (such as the
invariant domain  which combines both the previous domains), or to include
the Fermi and the free case (for which the natural domain is the linear
span of the $n$-particle vectors), we need the inequalities of [AcFaQu90].

Using the inequality (8.2) one can complete the simple stochastic integrals
in the natural seminorms given by the right hand side of (8.2) and
obtain the (vector) space of stochastic integrals (with respect to $X$).

The basic example of the situation described above is the following: let
\bigskip
\item{--} $ {\cal H} = \Gamma (
L^2( {\bf R}_+ ) $  the Fock space over the one-particle
   space $ L^2( {\bf R}_+ ) $
\item{--} $ {\cal D}= \{ \psi (f) = \sum_{n\geq 0} {\otimes ^n f  \over n! }:
\ f \in  L^2( {\bf R}_+ ) \} $ the set of exponential
   vectors in $ {\cal H}$.
\item{--} $\Phi = \psi (0) $ the vacuum state in $ {\cal H }$
\item{--} $ {\cal H }_{t]} = \Gamma ( \chi _{[0,t]} ) {\cal H } =
\Gamma ( L^2([0,t] ) \otimes  \Phi _{[t}$
where $t\geq  0 $ and    $\Gamma ( \chi _{[0,t]} ) $  is the
orthogonal projector defined by
$$\Gamma ( \chi _{[0,t]} ) \psi (f) = \psi ( \chi _{[0,t]} f )$$
\item{--} $A \ $ the annihilation field defined, on $ {\cal D}$ by the
relation:
$$A(f) \psi (g) = < f,g> \psi (g) $$
\item{--} $A^+$ the corresponding creation field, the adjoint of the
annihilation.

For all $ f\in L_{ \rm loc } ^2( {\bf R}_+)$, the additive processes
$ A( \chi _{(s,t]}f ) \ , \ A^+( \chi _{(s,t]}f)$
are regular semimartingales for the set  $ {\cal D}$ and, on this
domain, they satisfy the commutation relations
$$  [A(f), A^+(g)] = < f, g> \eqno (8.3)$$

The process
$ (A_t  , A^+_t )= (A( \chi _{(0,t]}) \ , \ A^+( \chi _{(0,t]}) $ is called
{\it the standard Boson Fock Brownian motion }.
In physics one uses more often the notation:
$$A(f)=\int_{\bf R} f(s) a_sd s \eqno(8.4)$$
where $a_s$ is an operator valued distribution (annihilation density)
satisfying the commutation relations
$$ [a_s,a^+_t]=\delta(t-s) \eqno(8.5)$$
The (distribution valued) process $(a_s,a^+_t)$ is called {\it quantum
white noise}.

The classical Brownian
motion is the process $ Q_t = A_t  + A^+_t $ (notice the continuum
analogue of the master field (1.9a)). The process
$$P_t:={1\over 2i}\{A(\chi_{[0,t]})-A^+(\chi_{[0,t]})\}\eqno(8.6)$$
is also a classical Brownian motion and one has
$$ [P_s,Q_t]=i \min\{s,t\}\  $$
in this sense one says that a quantum BM is a pair of non commuting
classical BM.

The space $H= \Gamma ( L^2(R))$ or  $(\Gamma ( L^2([0,+\infty ))))$
is the simplest model for the state space of a
quantum noise in the theory
of dissipative quantum phenomena (cf. [Ac90]).  It is canonically
isomorphic to the $L^2$-space of the increments of the Wiener
process (the white noise space) and its factorizing property
$$ \Gamma (L^2(R))   \cong \Gamma (L^2((- \infty ,t]))  \otimes
\Gamma (L^2((t,+\infty  )))\eqno(8.7)$$
corresponds to the independence of the increments of the Wiener
process.

The essence of Ito's formula is contained in the commutation relations (6.6)
and in the algebraic identity ( )

To show this, consider the increments of the creation and annihilation
operators processes:
$$A_t - A_s= A(\chi _{[s,t]}) \qquad ; \qquad A^+_t - A^+_s =
A^+(\chi _{[s,t]})$$
where $\chi _{[s,t]}(r) $ denotes the characteristic function of the
interval $[s,t] \subseteq {\bf  R } $ i.e.
$$\chi _{[s,t]} = 0 \qquad {\rm if } \ r \notin [s,t] \qquad ; \qquad
= 1 \ \ {\rm otherwise } $$
Now choose a small interval $[t,t+d t]$ and apply formula (8.6) above to
the increment of the annihilation process over this interval, i.e.
$$ d A_t=A_{t+d t}-A_t=A(\chi_{[t,t+d t]}) $$
Then for any continuous function $g$ one finds the approximation:
$$ d A_t\psi(g)=\langle\chi_{[t,t+d t]},g\rangle\psi(g)
=\left(\int_t^{t+dt}g(s)d s\right)\psi(g)\equiv g(t)d t\cdot\psi(g) $$
where for a {\bf  numerical } function $F(t)$ we write
$$F(t+d t)-F(t) \equiv 0 \Leftrightarrow
\lim_{dt \to 0} \bigl( sup_{ t \in [o,T]} (F(t+d t)-F(t)\bigr)/dt \to 0 \ ;
\ \forall T < + \infty $$
in this case we say that the difference $F(t+d t)-F(t)$ {\it is of
order\/} $o(dt)$.
This suggests that the topology of weak convergence on the set of the
exponential vectors is a natural one to give a meaning to a quantum analogue
of the Ito table. In fact the relation
$$d A_t\psi(g)\equiv g(t)d t\cdot\psi(g)$$
implies that the two-parameter families
$$A^2(\chi _{[s,t]}) \qquad , \qquad A^{+2}(\chi _{[s,t]})
\qquad , \qquad A^+(\chi _{[s,t]})\cdot A(\chi_{[s,t]})$$
are of order $o(dt)$ for the topology of weak convergence on the
exponential vectors $\psi(f)$
with $f$ continuous and square integrable.  In fact, for any pair
$\psi (f)$, $\psi (g)$ of such vectors one has,
$$< \psi (f), A^2(\chi _{[t,t+dt]}) \psi (g)> =
\biggl(\int_t^{t+dt} gds\biggr) ^2 <  \psi (f),  \psi (g)>\eqno(8.8a)$$
$$< \psi (f), A^{+2}(\chi _{[t,t+dt]}) \psi (g)> =
\biggl(\int_t^{t+dt} gds  \biggr) ^2 <  \psi (f),  \psi (g)>
\eqno(8.8b)$$
$$
\langle\psi_f,d A^+_td A_t\psi_g\rangle=\langle d A_t\psi_f,
d A_t\psi_g\rangle
\equiv (f(t)d t)^*(g(t)d t)\cdot\langle\psi_f,\psi_g\rangle\equiv
O(d t^2)\langle\psi_f,\psi_g\rangle $$
which are all of order $o(dt)$. Moreover
$$
< \psi (f), A^{+}(\chi _{[t,t+dt]}) \cdot A (\chi _{[t,t+dt]}) \psi (g)>
=\biggl(\int_t^{t+dt} {\overline f} gds \biggr) \cdot
<  \psi (f),  \psi (g)> \eqno(8.8c) $$
which is of order $dt$. Hence, in our notations:
$$d A^+_td A_t\equiv 0\ \qquad ; \ \qquad d A^+_td A^+_t\equiv dt $$
Similarly one proves that:
$$(d A_t)^2\equiv(d A_t^+)^2\equiv 0\ $$
However $ A(\chi _{[t,t+dt]}) \cdot A^+(\chi _{[t,t+dt]})$ is not of order
$o(dt)$ in the same topology since, from the Heisenberg commutation relation:
$$[A(f),A^+(g)] = <f,g>\eqno(8.9)$$
and from (8.8c), one deduces that
$$< \psi (f), A(\chi _{[t,t+dt]}) \cdot A ^{+}(\chi _{[t,t+dt]}) \psi (g)> =$$
$$= \biggl(\int_t^{t+dt} {\overline f}(s) g(s)ds \biggr) \cdot
<\psi (f),\psi (g)> + o(dt) \equiv {\overline f} (t)g(t)dt  \cdot
<\psi (f),\psi (g)> $$
Hence, in our notations:
$$d A_td A^+_t\equiv d t\ $$
The classical Ito table, deduced as an application of the above, is the
set of equations ( ), ( ), ( ) plus the obvious equation $dtdt \equiv 0$.
\bigskip \bigskip
Define $W_t=A_t+A^+_t$. If $s<t$
$$[W_s,W_t]=[W_s,W_s]+[W_s,W_{[s,t]}]=0$$
where, in the last identity, we have used the property that {\it the future
commutes with the past}.
%(an immediate consequence of (4.2)).
We conclude that $(W_t)$ is a commutative family and therefore $\{(W_t),
\Phi\}$ is a classical stochastic process. Moreover $(W_t)$ is mean 0 and
Gaussian since both $A_t$, $A^+_t$ are. Finally
$$d W^2_t=(d A_t+d A^+_t)^2
=d A^2_t+d A_td A^+_t+d A^+_td A_t+(d A^+_t)^2\equiv d t$$
Therefore $(W_t)$ is a classical Wiener process.

Another important example of stochastic process is the number (or gauge)
process.

Let be given a pre-Hilbert space $H$; let
$\Gamma(H)$ be the Fock space over $H$,
%$$(\otimes_{\hbox{symm. }}H)^0=\hbox{\bf C }\Phi:\qquad
$\Phi$ the Fock vacuum. For each $X\in B(H)$ with the norm less than 1,
define a bounded operator $\Gamma(X)$ on $\Gamma(H)$ by the relation
$$\Gamma(X)\psi(f):=\psi(Xf)\eqno(8.10)$$
A special case of (8.10) is $X=e^{ith}$ with the operator $h$ being
self-adjoint. By the Stone-von Neumann Theorem and (8.10), $(e^{ith})$, as well
as
$(\Gamma (e^{ith}))$, are unitary group on $H$, and $\Gamma (H)$ respectively.
Denote $N(h)$ the generator of $(\Gamma (e^{ith}))$, i.e.
%\psi(f)=\Phi+\sum^\infty_{n=1}{(\otimes f)^n\over\sqrt{n!}}&=
%\quad\hbox{exponential vectors}\cr
%e^{\Vert f\Vert^{2/2}}\psi(f)&\quad\hbox{coherent vectors}\cr}$$
$$e^{it N(h)}\psi(f):=\psi(e^{ith}f)=\Gamma(e^{ith})\psi(f)\eqno(8.11)$$
%$N(h)$ is well defined for $h=h^*$.
and moreover for an arbitrary operator $k$, we define $N(k)$ by complex
linearity:
$$N(k)=N\left(\left[{k+k^*\over2}\right]+i\left[{k-k^*\over2i}\right]\right):=
N\left({k+k^*\over2}\right)+iN\left({k-k^*\over2i}\right)\eqno(8.12)$$
The distribution of $N(h)$ in a coherent state (described by normalized
exponential vectors--coherent vectors)
is (a linear combination of) Poisson, i.e., if $h=h^*$
$$\eqalign{
\langle&e^{-{1\over2}\Vert f\Vert}\psi(f),\;e^{it N(h)}\cdot e^{-{1\over2}
\Vert f\Vert}\psi(f)\rangle=\cr
&=e^{-\Vert f\Vert^2}\langle\psi(f),\;\Gamma(e^{ith})\psi(f)\rangle=
e^{-\Vert f\Vert^2}\langle\psi(f),\psi(e^{ith}f)\rangle\cr
&=\exp\{-\Vert f\Vert^2+\langle f,e^{ith}f\rangle\}=\exp\langle f,
(e^{ith}-1)f\rangle\ \cr}\eqno(8.13)$$

{\it Example } Let $h=h^2=h^*$ be an orthogonal projection. Then:
$$e^{ith}-1=e^{it}\cdot h+h^\perp-1=(e^{it}-1)h\eqno(8.14)$$
therefore the
characteristic function of $N(h)$ is $\exp(e^{it}-1)\Vert hf\Vert^2$
which corresponds to a Poisson distribution with intensity
$$\Vert hf\Vert^2$$\bigskip\bigskip\bigskip\bigskip
\bigskip

\noindent{\bf (9.) Stochastic  limits and anisotropic asymptotics}
\bigskip

 Let us study the behaviour of correlation
functions when only some components of coordinates in a preferable frame
are suppose large or small. One of the most important examples
corresponds to the (2+2)-decomposition and it describes the case
when all longitudinal components in the central mass frame are assumed
much smaller then transersal ones.
Let $x^{\mu}$ be coordinates in the 4-dimensional Minkowski space-time
and  denote
$x^{\mu}=(y^{\alpha},z^{i})$,
$\alpha =0,1$, $i=2,3$. 2+2 Anisotropic asymptotics
of  Green functions of scalar selfinteracting theory
describe the behaviour of
the following correlation functions
$$G_{n}(\{\lambda y_{i},z_{i}\})=<\phi (\lambda y_{1},{z}_{1})
\phi (\lambda y_{2},{z}_{2})...
\phi (\lambda y_{n},{z}_{n})> \eqno(9.1)
$$ for $\lambda \to 0$.

For the scalar selfinteracting theory
 (9.1) is given by
$$<\phi (\lambda y_{1},{z}_{1})
\phi (\lambda y_{2},{z}_{2})...
\phi (\lambda y_{n},{z}_{n})>=\int \phi (\lambda
y_{1},{z}_{1})
\phi (\lambda y_{2},{z}_{2})...
\phi (\lambda y_{n},{z}_{n})\cdot   \eqno(9.2)
$$
$$\exp \{\int d^{4}x [{1\over 2}(\partial _{\alpha} \phi )^{2}
+(\partial _{i} \phi)^{2} +V(\phi )]
\} d\phi .
$$ Performing  in (9.2) the rescaling
$$\phi (\lambda y,z)=\tilde {\phi }(y,z)
$$
and the change of variables in the action $y\to \lambda y$ one gets
$$G_{n}(\{\lambda y_{i},z_{i}\})
=\int \tilde {\phi} (y_{1},{z}_{1})
\tilde {\phi} (y_{2},{z}_{2})...
\tilde {\phi} ( y_{n},{z}_{n})\cdot
$$
$$\exp \{\int d^{4}x {1\over 2}(\partial _{\alpha}
\tilde {\phi })^{2}
+\lambda ^{2}(\partial _{i}
\tilde {\phi})^{2} +\lambda ^{2}V(\tilde {\phi })
\} d\tilde {\phi},
$$
i.e. the theory with the effective action
$$
{\cal L}_{eff}={1\over 2}(\partial _{\alpha}\tilde { \phi})^{2}
+{\lambda ^{2}\over 2}(\partial _{i}
\tilde { \phi})^{2} +\lambda ^{2}
V(\tilde {\phi}) \eqno(9.3)
$$

Now we consider the asymptotics of the Green functions for the theory with
the action (9.3) when $\lambda \to 0$. Let us discuss  the free
action.
The free propagator for the action (9.3) has the form
$$
G_{\lambda }(y,z)=
{1\over {(2\pi )^{4}}}\int  {e^{ikx}\over
{k_{\alpha}^{2}+\lambda^{2}k_{i}^{2}}}dk
={1\over {(2\pi )^{2}}}{1\over {\lambda^{2}
y^{2}+z^{2}}} \eqno(9.4)
$$
Let us examine the limit of (9.4) for $\lambda \to 0$ in the sense
of theory of  distributions, i.e. the asymptotic behaviour of
the integral
$$
( G_{\lambda},f)={1\over {(2\pi)^{2}}}\int d^{2}yd^{2}z
{f (y,z)\over {\lambda^{2}y^{2}+ z^{2}}}
$$
when $\lambda \to 0$. Here  $f(y,z)$ is a test function.
One  gets
$$ G_{\lambda}(y,z)=
{1\over {4\pi}}  \delta^{(2)}(z)\ln {1\over {\lambda^{2}}}
+{1\over {4\pi^{2}}}{1\over {z^{2}}}
+ {1\over {4\pi}}\delta^{(2)}(z)\ln {1\over {y^{2}}}
 + o(1)]
$$
Here $$ {1\over {z^{2}}}= Reg{1\over {z^{2}}},$$
$$({1\over {z^{2}}}, f)=\int_{|z|\leq 1} d^{2}z
{f (z)-f (0)\over {z^{2}}}+
\int d^{2}y\int_{|z|> 1}d^{2}z
{f(z)\over {z^{2}}}
$$
For a consideration of anisotropic asymptotics in scalar and gauge
theories see [AV94].
\bigskip

\noindent{\bf (10.)  Stochastic bosonization}
\bigskip

Let $\Gamma_-({\cal H}_1)$ denote the Fermi Fock space on the
1--particle space ${\cal H}_1$ and let, for $f\in{\cal H}_1$, $A(f)$,
$A^+(f)$ denote the creation and annihilation operators on
$\Gamma_-({\cal H}_1)$ which satisfy the usual canonical anticommutation
relations (CAR):
$$A(f)A^+(g)+A^+(g)A(f)=\langle f,g\rangle$$
The main idea of stochastic bosonization is that, in a limit to be specified
below, two Fermion operators
give rise to a Boson operator. To substantiate the idea
let us introduce the operators
$${\cal A}(f,g):=A(g) A(f),\qquad {\cal A}^+(f,g) := \bigl({\cal A}(f,g)
\bigr)^*$$
then by the CAR we have that
$${\cal A}(f,g){\cal A}^+(f',g')=A(g) A(f)A^+(f') A^+(g')=$$
$$=<g,g'><f,f'>-<f,f'><g,g'>+{\cal A}^+(f',g'){\cal A}(f,g)+R(f,g;f',g')
$$
i.e.
$$[{\cal A}(f,g),{\cal A}^+(f',g') ]= <(f\otimes g)
(f'\otimes g')>+R(f,g;f',g')
$$
where we introduce the notations
$$R(f,g;g',f'):=<f,g'>A^+(f')A(g) -<f,f'>A^+(g')A(g)
-<g,g'>A^+(f')A(f)$$
$$<(f\otimes g)(f'\otimes g')> :=<f,f'><g,g'>-<f,g'><g,f'> $$
Moreover,
$${\cal A}(f,g){\cal A}(f',g')  ={\cal A}(f',g'){\cal A}(f,g)
$$

One can prove that, in the stochastic limit, the
remainder term $R$ tends to zero so that, in this limit, the {\it
quasi CCR\/} become {\it bona fide\/} CCR.

The first step in the stochastic limit of quantum field theory is to
introduce the {\it collective operators\/}. In our case we associate to
the quadratic Fermion operator
the collective creation operator defined by:
$$A^+_{\lambda}(S,T;f_0,f_1):=\lambda\int_{S/\lambda^2} ^{T/
\lambda^2}
e^{i\omega t} A^+(S_tf_0)A^+(S_tf_1)dt\eqno(10.1)$$
where $\omega$ is a real number,
$S\le T$, $f_0,f_1\in {\cal K}$ and $S_t:{\cal H}_1\to{\cal H}_1$
is the one--particle dynamical
evolution whose second quantization gives the free evolution of the
Fermi fields, i.e.
$$A(f)\mapsto A(S_tf)\ ;\quad A^+(f)\mapsto A^+(S_tf)$$

The collective annihilation
operators are defined as the conjugate of the collective creators.

The introduction of operators such as the right hand side of (10.1) is a
standard technique in the stochastic limit of QFT, we refer to [AcAlFriLu]
for a survey and a detailed discussion. Here we limit ourselves to state
that the choice (10.1) is dictated by the application of first order
perturbation theory to an interaction Hamiltonian.

Having introduced the collective operators, the
next step of the stochastic approximation is to compute the
2--point function
$$<\Phi, A_\lambda(S,T;f_0,f_1) A^+_\lambda(S',T';f'_0,f'_1)
\Phi>=$$
$$=\lambda^2 \int_{S/\lambda^2} ^{T/\lambda^2}dt\int_{S'/\lambda^2}
^{T'/\lambda^2}dse^{i\omega (t-s)} <\Phi,A(S_tf_0)A(S_tf_1)
A^+(S_sf'_0)A^+(S_sf'_1)\Phi>\eqno(10.2)$$
By the CAR, the scalar product in the right hand side of (10.2) is equal to
$$<S_tf_0,S_sf'_1><S_tf_1,S_sf'_0> -<S_tf_0,S_sf'_0><S_tf_1,S_sf'_1>$$

By standard arguments [AcLu1] one proves that
the limit, as $\lambda\to 0$, of (10.2) is
$$<\chi_{[S,T]} ,\chi_{[S',T']}>_{L^2({\bf R})}\cdot
\int_{-\infty}^\infty ds e^{i\omega s}
\bigl[<f_0,S_sf'_1><f_1,S_sf'_0> -<f_0,S_sf'_0><f_1,S_sf'_1>
\bigr]$$
Let us introduce on the algebraic
tensor product ${\cal K}\odot{\cal K}$
the pre--scalar product $~~~~$ $(\cdot|\cdot)$ defined by:
$$(f_0\otimes f_1 |f'_0\otimes f'_1 ):=$$
$$=\int_{-\infty}^\infty ds e^{i\omega s}
\bigl[<f_0,S_sf'_1><f_1,S_sf'_0> -<f_0,S_sf'_0><f_1,S_sf'_1>
\bigr]$$
and denote by ${\cal K}\otimes_{\rm FB}{\cal K}$ the Hilbert space
obtained by completing ${\cal K}\odot{\cal K}$ with this scalar product.
Now let us consider the correlator
$$<\Phi, \prod_{k=1}^nA_\lambda^{\varepsilon
(k)}(S_k,T_k;f_{0,k},f_{1,k})\Phi>\eqno(10.3)$$
where,
$$A^\varepsilon:=\cases{A^+,\ &if $\varepsilon=1$,\cr
A,\ &if $\varepsilon=0$\cr} $$

By the CAR, it is easy to see that
(10.3) is equal to zero if the
number of creators is different from the number of annihilators, i.e.
$$\Bigl|\{k;\ \varepsilon (k)=1\}\Bigr|\not=
\Bigl|\{k;\ \varepsilon (k)=0\}\Bigr| $$
or if there exsits a $j=1,\cdots, n$ such that the number of creators on
the left of $j$ is greater than the number of annihilators with the same
property, i.e.
$$\Bigl|\{k\le j;\ \varepsilon (k)=1\}\Bigr|>
\Bigl|\{k\le j;\ \varepsilon (k)=0\}\Bigr| $$
\bigskip

$\underline{\rm {\bf THEOREM}\ (10.1)}$  The limit, as $\lambda\to 0$,
of (10.3) is equal to
$$<\Psi, \prod_{k=1}^na^{\varepsilon
(k)}(\chi_{[S_k,T_k]}\otimes f_{0,k}\otimes_{\rm
FB}f_{1,k})\Psi>$$
where, $a, a^+$ and $\Psi$
are (Boson) annihilation, creation operators on the Boson Fock space
$\Gamma_+(L^2({\bf R})\otimes {\cal K}\otimes_{\rm FB}{\cal K})$ respectively
and $\Psi$ is the vacuum vector.

For a consideration of the stochastic bosonization in a model with
an interaction see [AcLuVo94a].
\bigskip

\noindent{\bf (11.) The interacting Fock module and QED}

\bigskip
Here we briefly review how the stochastic limit for quantum
electrodynamics without dipole approximation naturally leads to the
non--crossing diagrams, corresponding to a nonlinear deformation of the
Wigner semicircle law, as well as to a {\it
generalization of the free algebra\/} and to the introduction of the
so--called {\it interacting Fock space\/} [AcLu93a] . For simplicity we
shall consider the simplest of this model: a {\bf polaron type
particle}, interacting via the minimal coupling with
a quantum EM field described by the Hamiltonian
$$H=H_S+H_R+H_I=H_0+H_I\eqno(11.1)$$
where denoting $p=(p_1,p_2,p_3)$ the momentum operator,
the particle (system) Hamiltonian is
$$H_S={p^2\over2}\,\otimes1\eqno(11.2)$$
where the reservoir Hamiltonian is
$$H_R=1\otimes\int_{{\bf R}^3}|k|a^*_ka_kdk\eqno(11.3)$$
$a_k$ and $a^*_k$ are bosonic annihilation
and creation operators and the interaction Hamiltonian is
$$\lambda H_I=\lambda\int_{{\bf R}^3}g(k)e^{ikq}p\otimes{a_k\over
\sqrt{|k|}}\,dk+\hbox{ h.c.}\eqno(11.4)$$
where $g(k)$ (11.4) is a cutoff test function, $\lambda$ is a coupling
constant and $p$ and $q$ satisfy the commutation relations $[q,p]=i$.

The Hamiltonian (11.1) is an operator in the tensor product of Hilbert
spaces of the particle and the field, ${\cal H}\otimes\Gamma({\cal H})$,
where ${\cal H}=L^2({\bf R}^3)$, $\Gamma({\cal H})$ is the bosonic
Fock space.

In the weak coupling limit (stochastic limit) one considers the
asymptotic behaviour for $\lambda\to0$, of the {\it time scaled\/}
evolution operator in the interaction
representation, $U(t/\lambda^2)$, where:
$$U(t)=e^{itH_0}e^{-itH}$$
satisfies the equation
$${d\over dt}\,U(t)=-i\lambda H_I(t)U(t)$$
and $H_I(t)$ is the interaction representation Hamiltonian:
$$H_I(t)=e^{itH_0}H_Ie^{-itH_0}=A^*(S_tg)(-ip)+\hbox{ h.c. }$$
$$A^+(S_tg)=\int_{{\bf R}^3}dke^{-ikq}e^{itkp}e^{-it|k|^2/2}(S_tg)(k)
\otimes a_k$$
The factor $\exp {-it|k|^2/2} $ shall be incorporated
into the free one particle evolution $S_t$ giving rise to the new
evolution $\exp {-it|k|^2/2}$ still denoted, for simplicity, with
the same symbol $S_t$.
One defines the {\it collective annihilator process\/} by
$$A_{\lambda}(S_1,T_1,g) =\lambda\int^{T_1/\lambda^2}_{S_1/\lambda^2}dt
\int_{{\bf R}^3}dke^{-itkp}e^{ikq}e^{-itk^2_2}\otimes\overline{S_tg}
(k)a_k\eqno(11.5)$$
and its conjugate, $A^*_\lambda(t)$, and consider the limit when
$\lambda\to0$. If one puts $\lambda=1$ $(N\to\infty)$ and replaces
$T/\lambda^2=NT$ by its integer part, the analogy between (12.5) and
(11.6a) below) and the functional central limit theorem of Section (7.)
becomes apparent.
Let $K\subset L^2({\bf R}^3)$ be the subspace of the
Schwartz functions such that for any pair $f,g\in K$ the condition
$$\int|\langle f,S_tg\rangle|dt<\infty$$
is satisfied.\bigskip

\noindent{\bf Theorem 11.1.} {\sl For any $n\in{\bf N}$ and test
functions $g_1,\dots,g_n$, from $K$ the limit of the collective scalar product
$$\lim_{\lambda\to 0} < A^*_\lambda(S_1,T_1;g_1)\dots
A^*_\lambda(S_n,T_n;g_n)\xi\otimes\Omega , \eta\otimes\Omega>\eqno(11.6a)$$
exists and is equal to
$$ \langle A^*(\chi_1\times g_1)\dots A^*(\chi_n\times
g_n)\xi\otimes\psi\,,\ \eta \otimes \psi\rangle\eqno(11.6b)$$
where $\chi_j=\chi_{[S_j,T_j]}$ is the characteristic function of the
interval $[S_j,T_j](\chi_j(t)=1$ if $t\in[S_j,T_j]$; $=0$ otherwise) and
the right hand side of (11.5) is an inner product in a limit Hilbert
spaces which is the tensor product of a Hilbert module with the system
Hilbert space, endowed with a new type of scalar product, which is
described in the following.}\bigskip

We shall now introduce the following notations.

For each $f\in K$ define
$$\tilde f(t):=\int_{{\bf R}^d}e^{-ik\cdot q}e^{ik\cdot p}(S_tf)(k)dk$$
then $\tilde f:{\bf R}\to{\cal B}(L^2({\bf R}^d))$.
Denote by ${\cal P}$ the momentum algebra of the particle (system) i.e.
the von Neumann--algebra generated by
$\{e^{ik\cdot p}:k\in{\bf R}^d\}$, and by ${\cal F}$ the
${\cal P}$--right--linear span of $\{\tilde f:f\in K\}$.
Then the tensor product $L^2({\bf R})\otimes{\cal F}$ is a
${\cal P}$ (in fact $1\otimes{\cal P}$)--right module, on which we introduce
the inner product
$$(\alpha\otimes\tilde f|\beta\otimes\tilde g):=\langle\alpha,\beta
\rangle_{L^2({\bf R})}\int_{\bf R}du\int_{\bf R}d_ke^{-iuk\cdot p}
\overline f(k)(S_ug)(k)\eqno(11.7)$$
which is a positive ${\cal P}$--right--sesquilinear form.

Starting from $L^2({\bf R})\otimes{\cal F}$ and taking the quotienting
by the elements
of zero norm with respect to the inner product $(\cdot|\cdot)$ given by
(11.7), one obtains a (pre--)Hilbert module, still denoted by
$L^2({\bf R})\otimes {\cal F}$.
By $(L^2({\bf R})\otimes{\cal F})^{\otimes n}$ we shall denote the $n$--th
algebraic tensor power of $L^2({\bf R})\otimes{\cal F}$,
on which we introduce the inner product
$$((\alpha_1\otimes\tilde f_1)\otimes\dots\otimes(\alpha_n\otimes\tilde
f_n)|(\beta_1\otimes\tilde g_1)\otimes\dots\otimes(\beta_n\otimes\tilde
g_n)):=$$
$$:=\prod^n_{h=1}\langle\alpha_n,\beta_n\rangle_{L^2({\bf R})}\cdot
\int_{{\bf R}^n}du_1\dots du_n\int_{{\bf R}^{nd}}dk_1\dots dk_n$$
$$\prod^n_{h=1}[e^{-iu_hk_h\cdot p}\overline f_h(k_h)(S_{u_h}g_h)(k_h)]
\cdot\exp\left(i\sum_{1\leq r\leq h\leq
n-1}u_rk_rk_{h+1}\right)\eqno(11.8)$$
Notice that because of the exponential factor, even if the $p$ in the
exponential were a scalar, and not, as it is an operator on the particle
space, the left hand side of (11.8) would not be equal to
$$(\alpha_1\otimes\tilde f_1|\beta_1\otimes\tilde
g_1)\cdot\dots\cdot(\alpha_n\otimes\tilde f_n|\beta_n\otimes\tilde
g_n)\eqno(11.9)$$
i.e. to the usual scalar product in the full Fock space.

Thus, with respect to the usual Fock (or full Fock) space, in the
expression (11.8) there are two new features:
\item{i)} the inner product is not scalar valued but takes values in the
momentum algebra of the system (i.e. particle) space.
\item{ii)} If we interpret $(L^2({\bf R})\otimes{\cal F})^{\otimes n}$ as
a kind of $n$--particle space, then the exponential factor in (11.8)
indicates that {\bf these $n$--particle interact}.

The feature (i) tells us that we are in presence of an Hilbert module
(over the momentum space of the particle). The feature (ii) is a remnant
of the interaction: before the limit the different modes of the field
interacted among themselves only through the mediation of the particle.
After the limit {\bf there is a true self--interaction}. Although
complex, formula (11.8)
seems to be the first one to describe in an explicit way a real
self--interaction between all the modes of a quantum field.

It would be therefore very interesting to obtain a {\bf relativistic
generalization} of formula (11.8), which should be an achievable goal
because the interaction among the modes occurs via the scalar product,
which is an invariant expression.

In analogy with the usual Fock space, we introduce the direct sum
$$\Gamma(L^2({\bf R})\otimes{\cal F}):=
{\bf C}\cdot\Psi\oplus\bigoplus^\infty_{n=1}(L^2({\bf R})\otimes{\cal
F})^{\otimes
n}\eqno(11.10)$$
where the unit vector $\Psi$ is called {\it the vacuum\/}. If we endow
each {\it $n$--particle space\/} in (11.10) with the inner product (11.8),
we do not obtain the usual full Fock space, but a new object that,
because of the reasons explained above, has been called {\bf the
interacting  Fock module}.

In order to pursue the analogy with the usual (full) Fock space,
define the creator by
$$A^+(\alpha\otimes\tilde f)[(\alpha_1\otimes\tilde f_1)\otimes\dots\otimes
(\alpha_n\otimes\tilde f_n)]:=
(\alpha\otimes\tilde f)\otimes(\alpha_1\otimes\tilde
f_1)\otimes\dots\otimes(\alpha_n\otimes\tilde f_n)$$
and the annihilator by
$$A(\alpha\otimes f):=[A^+(\alpha\otimes f)]^+$$
One easily shows that the action of the annihilator is the one suggested
by the obvious analogy with the Fock case, although now the inner
product is an operator\bigskip

\noindent{\bf Theorem 11.2}. {\sl
$$A(\alpha\otimes\tilde f)[A^+(\alpha_1\otimes\tilde f_1)\dots
A^+(\alpha_n\otimes\tilde f_n)\Psi]$$
$$=(\alpha\otimes f)|\alpha_1\otimes\tilde f_1)A^+(\alpha_2\otimes\tilde
f_2)\dots A^+(\alpha_n\otimes\tilde f_n)\Psi$$
where $\Psi$ is the vacuum of $\Gamma(L^2({\bf R})\otimes{\cal
F})$.}\bigskip

The explicit calculation of the quantities of physical interest in made
possible by the following:\bigskip

\noindent{\bf Theorem}. {\sl $\forall\,n\in{\bf N}$
$\varepsilon\in\{0,1\}^n$, $A^0:=A$, $A^1:=A^+$
$$\langle\Psi,A^{\varepsilon(1)}(\alpha_1\otimes\tilde f_1)\dots
A^{\varepsilon(n)}(\alpha_n\otimes\tilde f_n)\Psi\rangle\eqno(11.11)$$
is 0 if $\{\varepsilon(1),\dots,\varepsilon(n)\}$ does not
allow a non--crossing pair partitions; if
$\{\varepsilon(1),\dots,\varepsilon(n)\}$
allows non--crossing pair partition, then it is unique and $n=2j$ for
some $j$. In
this case if we denote by $1<m_1<m_2<\dots<m_j=2j=n$ the position of
the creators and by $\{m'_h\}^j_{h=1}$ the position of the corresponding
annihilators, in the sense that
the non--crossing pair partition is given by:
$$(m'_1,m_1),\ (m'_2,m_2),\dots,(m'_j,m_j)$$
then (12.11) is given by
$$\prod^j_{h=1}\langle\alpha_{m_h},\alpha_{m'_h}\rangle_{L^2({\bf R})}
\int_{\bf R}du_1\dots\int_{\bf R}du_j\int_{{\bf R}^{jd}}dk_1\dots
dk_j$$
$$\left[\prod^j_{h=1}(Su_hfm_h)(k_h)\overline f_{m'_h}(k_h)\cdot
e^{iu_hk_h\cdot p}\right]\cdot$$
$$\cdot\exp\left(i\sum^{j-1}_{h=1}\sum^j_{r=h+1}u_hk_h\cdot
h_r\chi_{(m'_r,m_r)}(m_h)\right)$$}\bigskip

It can also be proved
that the limit (in the sense of quantum convergence in law) of
the rescaled wave operator
$U_{t/\lambda^2}$ exists
and satisfies a quantum stochastic differential equation on $\Gamma(L^2
({\bf R})\otimes{\cal F})\otimes{\cal
H}_0$ (cf. the equation (11.2) of [AcLu]).\medskip

We shall not discuss here the quantum stochastic differential equation,
associated to this model, because it requires the notion of {\it
stochastic integration over Hilbert modules\/}, which introduces several
new features with respect to the usual quantum stochastic integration.
For this we refer to the paper [AcLu91], [AcLu92], [AcLu93a], [AcLu93b],
for the model we are discussing
here and to [Lu92a], [Lu92b], [Lu94], for the general theory.\bigskip\medskip
\centerline{\bf References}\bigskip

\bigskip
[AFL82]  Accardi L., Frigerio A., Lewis J.:
Quantum  stochastic  processes
Publications  of the Research institute for  Mathematical
Sciences Kyoto University 18 (1982) 97-133.

[Ac90] Accardi L.: An outline of quantum probability.
Unpublished manuscript. (1990)
%  Review on Q. Prob.

[St] Stanley H.E.: Phase transitions and critical phenomena.
Oxford: Cambrige University Press 1971

[tHo74] 't Hooft G.: Nucl.Phys. B72 (1974) 461

[BrZ] Brezin E. and Zinn-Justin J.:Phys.Rev. D14 (1976) 2615

[Ar76] I.Ya. Aref'eva: Theor. Math. Phys. 29 (1976) 147;
                       Ann. Phys. 117 (1979)

[BIPZ] E. Brezin, C. Itzykson, G.Parisi and J.-B. Zuber,
Comm.Math.Phys 59(1978)35

[Wit] E. Witten , in Resent Developments in Gauge Theories ,
eds. G. 'tHooft et. al. Plenum Press,New York and London (1980)

[MM] Yu. Makeenko and A. A. Migdal, Nucl. Phys. B188 (1981) 269

[Ha80] O. Haan, Z. Physik C6 (1980) 345

[Sin94] I. Singer,Talk at the Congress of Mathematical Physics, Paris (1994)

[Ar81] I.Ya. Aref'eva, Phys. Lett. 104B (1981) 453.

[AV91] I.Ya. Aref'eva and I. V. Volovich Phys. Lett. B 268 (1991) 179

[Do94] M. R. Douglas, "Large N Gauge Theory - Expansions and Transitions,"
May 1994 lectures at the ICTP, hep-th/9409098

[GG94] R. Gopakumar and D. Gross, Princeton preprint PUPT-1520,hep-th/9411021

[AV94] I.Ya. Aref'eva and I. V. Volovich, Anisotropic
Asymptotics and High Energy Scattering, SMI-13-94, hep-th/9412155.

%[Are94] I.Ya. Aref'eva: The Large $N$ Expansion in Quantum Field Theory and
%Statistical Mechanics, eds. E. Brezin and S. Wadia, World Sci, 1994.

[APVV] I.Ya. Aref'eva, K.Parthasaraty, K.Viswanathan and
I.V.Volovich,  Mod.Phys.Lett.A9 (1994)689

[Sch\"u93] Sch\"urmann, M.: White noise on bialgebras. (Lect.
Notes Math., vol. 1544). Berlin Heidelberg New York: Springer 1993

[Sch\"u94] Sch\"urmann, M.: Direct sums of tensor products and
non-commutative independence. To appear in J. Funct. Anal.

[Sp90] Speicher R.:
A new example of independence and white noise,
Probab. Th. Rel. Fields 84 (1990), 141--159.

[K\"uSpe90] Speicher R., K\"ummerer B.:
Stochastic intergration on the Cuntz algebra $0_\infty$.
Func. Anal. Acad. Press. Vol. 103 No.2. 1992, preprint (1990).

[SpevW92] Speicher R., Von Waldenfels W.: A General Central limit theorem and
invariance principle.
in: Quantum Probability and Related Topics, World Scientific, OP--PQ IX (1994)
Preprint 707 1992.

[BoSp91] Speicher R., M.Bozejko: $\psi$-independent and symmetrized white
noises.
in: Quantum Probability and Related Topics, World Scientific, OP--PQ VI (1991)

[Sum90] Summers S.J.:
On the independence of local algebras in quantum field theory,
Rev. Math. Phys. 2 (1990), 201--247.

[vWI88] von Waldenfels W., Ion P.D.F.:
An algebraic approach to quantum stochastics. Unpublished manuscript (1988)

[GivWa78] von Waldenfels W., Giri N.:
An Algebraic Version of the Central Limit Theorem.
Z. Wahrscheinlichkeitstheorie verw.  Gebiete 42, 129--134 (1978).

[vWa78] von Waldenfels, W.:
An algebraic central limit theorem in the anticommuting case.
Z. Wahrscheinlichkeitstheorie verw.  Gebiete 42, 135--140 (1978).

[Hu71] Hudson R.L., Cushen C.D.
A Quantum-Mechanical Central Limit Theorem.
Journ. of Appl. Probab. 8.3(1971)454-469

[Hu73] Hudson R.L.
A Quantum-Mechanical Central Limit Theorem for
 anti-commuting Observables,
 Journ.of Appl.Probab. 10.3(1973)502=509

[Fag90]
Fagnola F.
Quantum stochastic integration with respect to free noises
 Preprint Volterra N. 37 (1990),
in: Quantum Probability and Related Topics, World Scientific, OP--PQ VI (1991)

[AAV93] L. Accardi, I.Ya. Aref'eva and I.V. Volovich, unpublished notes (1993).

%[Sin94] I. Singer, Talk at the Congress of Mathematical Physics, Paris, 1994.

[Voi92] Voiculescu D., Dykema K.J., Nica A.:
Free random variables. CRM Monograph Series, Vol. 1, American Math. Soc. (1992)

[Voi91] D.V. Voiculescu, Invent. Math., 104 (1991) 201--220.

%[Wig] Wigner E.P.: Characteristic vectors of bordered matrices withh

[AcBa87] Accardi L.,  A. Bach:
The harmonic oscillator as quantum central limit of Bernoulli processes.
accepted by : Prob. Th. and Rel. Fields, Volterra preprint (1987)

[Lu89] The Boson and Fermion quantum Brownian motions as quantum central limits
of quantum Bernoulli processes. Bollettino U.M.I., (7) 6--B (1992),
Volterra preprint (1989)

[AcFaQu90] Accardi L., Fagnola F., Quaegebeur
Quantum Stochastic Calculus
Journ. Funct. Anal. 104 (1992) 149--197
Volterra preprint N. 18 (1990)

[Par93] Parthasarathy K.R.:
Quantum Stochastic Calculus.
Birkheuser (1993)

[Fag92] Fagnola F.:
On the GNS representation of a free $C^*$-algebra.
Preprint,1991, Rendiconti Accademia dei Lincei (1992)

[AcLu92] Accardi L., Lu Y.G.:
The Wigner Semi--circle Law in Quantum Electro Dynamics.
submitted to: Commun. Math. Phys.  (November 1992)
Volterra preprint N.126 (1992)

[Ac90] Accardi L.:
Noise and dissipation in quantum theory.
Reviews in Math. Phys. 2(1990) 127-176

[Verb89] Verbeure A., Goderis D., Vets P.:
Non--commutative Central Limits,
Probability Theor. Rel. Fields 82, pp. 527-544, 1989a

[AcLu90a] Accardi L., Lu Y.G.:
 The low density limit of quantum systems,
 Jour. Phys. {\bf A}: 24(1991)3483--3512

[AcLu90b] Accardi L., Lu Y.G.:
Quantum central limit theorems for weakly dependent maps (I), (II),
Acta. Math. Hungar. (1994), Volterra preprint N.54 (1990)

[HuPa84a] Hudson R.L., Parthasarathy K.R.
Construction of quantum diffusions.
in : Quantum Probability and applications to the quantum theory of
irreversible processes, L.Accardi, A.Frigerio, V. Gorini (eds.)
Springer LNM N 1055 (1984)

[HuPa84b] Hudson R.L., Parthasarathy K.R.  Quantum Ito' s formula and
stochastic
evolutions.
Comm. Math. Phys. 93(1984)301-323

[AcLuVo93] L. Accardi, Y.G.Lu, I. Volovich:
The Stochastic Sector of Quantum Field Theory.
Volterra Preprint N.138, 1993;
to appear in: Matematicheskie Zametki (1994)

[AcLu91] Accardi L., Lu Y.G.:
 From Markovian approximation  to a new type of quantum stochastic calculus.
in: Quantum Probability and Related Topics, World Scientific, OP--PQ VII (1993)
Volterra preprint N.68 (1991)

[AcLuVo94a] Accardi L., Lu Y.G., Volovich I.:
On the stochastic limit of Quantum Chromodynamics
in: Quantum Probability and Related Topics, World Scientific, OP--PQ IX (1994)
Volterra preprint (1994)

[AcLu93a] L. Accardi, Y.G. Lu: Wiener noise versus Wigner noise in
quantum electrodynamics.
in: Quantum Probability and Related Topics, World Scientific, OP--PQ VIII
(1993)
Volterra preprint N. 131 (1993)

[AcLu93b] Accardi L.,  Lu Y.G.:
Quantum Electro Dynamics: the master and the Langevin equation.
{\it Mathematical Approach to Fluctuations }
% talk Kyoto International Institute for Advanced Studies, May 18-22, 1992
T.Hida (ed.) World Scientific (1994)
Volterra preprint N.133 (1993)

[AcLuVo94b] L. Accardi, Y.G. Lu, I. Volovich,
Stochastic bosonization in arbitrary dimensions,
Preprint Volterra (1994)

[AcLuMa94] L.Accardi, Y.G.Lu, V.Mastropietro
Stochastic bosonization for a $d=3$ Fermi system
Preprint Volterra (1994)

[Lu92a] Lu Y.G.
Quantum stochastic calculus on Hilbert module,
Math. Z. (1993), Volterra preprint N. 106, (1992)

[Lu92b] Lu Y.G.
Quantum stochastic calculus
The passage from quantum system with continuous spectrum to quantum
Poisson processes on Hilbert module,
Volterra preprint N. 106, (1992)

[Lu94]  Lu Y.G.:
A note on free stochastic calculus on Hilbert modules and its
applications,
Volterra Preprint n. 186, october 1994

[Wil94] K.G. Wilson, T.S. Walhout,
A. Harimdranas, Y. Min Zhang, R.J. Perry, S.
Glazek. Non perturbative QCD: a weak coupling treatment on the light
front. Preprint April 1994.

[Wig] Wigner E.P.: Characteristic vectors of bordered matrices with
infinite dimensions, Ann. of Math. 62(1955) 548--564

\bye